\documentclass[12pt]{article}

\usepackage{cite}
\usepackage{epsfig}
\usepackage{amsmath,amssymb}
\usepackage{tabularx}
\usepackage{pstricks}
\usepackage{array}

\def\mathswitch#1{\relax\ifmmode#1\else$#1$\fi}
\def\mathswitchr#1{\relax\ifmmode{\mathrm{#1}}\else$\mathrm{#1}$\fi}

\newcommand{\tev}{\,\, \mathrm{TeV}}
\newcommand{\gev}{\,\, \mathrm{GeV}}

\newcommand{\SLASH}[2]{\makebox[#2ex][l]{$#1$}/}

\newcommand{\Eslash}{\SLASH{E}{.3}\,}

\newcommand{\anc}{\rule{0mm}{0mm}}

\newcommand{\mycaption}[1]{\caption{\sl #1}}

\begin{document}
\thispagestyle{empty}

\def\thefootnote{\fnsymbol{footnote}}

\begin{flushright}
PITT-PACC-1406 \\
\end{flushright}

\vspace{1cm}

\begin{center}

{\Large\sc {\bf Heavy Color-Octet Particles at the LHC}} 
\\[3.5em]
{
Chien-Yi~Chen$^1$, Ayres~Freitas$^2$, Tao~Han$^2$, and Keith~S.~M.~Lee$^{3,4}$
}

\vspace*{1cm}
 
{\sl $^1$ Department of Physics, Brookhaven National Laboratory, Upton, NY 11973, U.S.A.
\\[1em]
\sl $^2$
PITTsburgh Particle physics, Astrophysics, and Cosmology Center (PITT PACC),\\
Department of Physics \& Astronomy, Univ.~of Pittsburgh,
Pittsburgh, PA 15260, 
U.S.A.
\\[1em]
\sl $^3$
Perimeter Institute for Theoretical Physics,  Waterloo, ON N2L$\,$2Y5, Canada
\\[1em]
\sl $^4$
Department of Physics \& Astronomy, Univ.~of Waterloo, Waterloo, ON N2L$\,$3G1, Canada
}

\end{center}

\vspace*{2.5cm}

\begin{abstract}

Many new-physics models, especially those with a color-triplet top-quark 
partner, contain a heavy color-octet state. The ``naturalness'' argument 
for a light Higgs boson requires that the color-octet state be not much heavier 
than a TeV, and thus it can be pair-produced with large cross sections at 
high-energy hadron colliders. 
It may decay preferentially to a top quark plus a top partner, which 
subsequently decays to a top quark plus a color-singlet state. This singlet 
can serve as a WIMP dark-matter candidate. Such decay chains lead to a 
spectacular signal of four top quarks plus missing energy. 
We pursue a general categorization of the color-octet states and their decay 
products according to their spin and gauge quantum numbers. We review the 
current bounds on the new states at the LHC and study the expected discovery 
reach at the 8-TeV and 14-TeV runs. We also present the production rates at a 
future 100-TeV hadron collider, where the cross sections will be many orders
of magnitude greater than at the 14-TeV LHC.
Furthermore, we explore the extent to which one can determine the color octet's 
mass, spin, and chiral couplings.
Finally, we propose a test to determine whether the fermionic color octet 
is a Majorana particle. 

\end{abstract}

\setcounter{page}{0}
\setcounter{footnote}{0}

\newpage


\section{Introduction}

The historic discovery of the Higgs boson has led particle physics to an 
interesting juncture. On the one hand, for the first time in history, we 
have a consistent relativistic quantum-mechanical model, the Standard Model 
(SM), that is valid all the way up to the Planck scale. On the other hand, 
there remain many unanswered theoretical and observational questions, which 
imply the need for physics beyond the SM. The putative ``naturalness'' 
of a light Higgs boson is arguably a strong indication of new physics near 
the TeV scale, and a top-quark partner is eagerly anticipated as a 
cure for the quadratic sensitivity of the Higgs mass to the new-physics scale.

Besides the color-triplet top-quark partner, many new-physics models  
contain a heavy color-octet state. The naturalness argument 
requires the color-octet state to be not much heavier than the TeV scale 
\cite{Bardimo}, which should be accessible at 
LHC energies (for a recent account, see for example Ref.~\cite{Papucci:2011wy} 
and references therein). 
Examples of electrically neutral 
color-octet particles include the gluino in supersymmetry \cite{susy}, 
techni-rhos \cite{tc} or top-gluons \cite{topgl} in models with strong 
TeV-scale dynamics, and Kaluza-Klein (KK) gluons in models with universal 
extra dimensions \cite{ued}. 
For large regions of parameter space in these models,
the color-octet particles decay preferentially to a top quark plus a heavy
top-quark partner, either owing to large couplings between the color
octet and the top-quark partner or because other new particles are very 
massive and thus effectively decoupled. The top partner subsequently decays 
to a top quark plus a color-singlet state. 
These decay chains lead to a spectacular signal of four top 
quarks plus missing energy. 

In this paper, we model-independently study processes of the form
\begin{equation}
pp \to {\cal Z\bar{Z}} \to t\bar{t}\ Y\bar{Y} \to t\bar{t}t\bar{t} \, X\bar{X} \,,
\label{eq:process}
\end{equation}
where ${\cal Z}$ is a new color-octet particle, $Y$
a new color-triplet particle (top partner), and $X$ a color singlet. The
electrically neutral $X$ is assumed to be stable and thus could be a dark-matter
candidate, which would manifest itself as missing energy in a collider
experiment. For all new particles in this process ($X$, $Y$ and ${\cal Z}$), we
consider different spin assignments (0, 1/2 and 1).
We also distinguish the possibility that the color octet may or may not be 
its own antiparticle (${\cal Z=\bar{Z}}$ or ${\cal Z\neq\bar{Z}}$).
Each combination is exemplified by particles in
well-motivated new-physics models. For example, ${\cal Z}$ could be the gluino, $Y$ a scalar top, and $X$ the 
lightest neutralino in the Minimal Supersymmetric Standard Model (MSSM). 
This case has been studied extensively (see, for example, 
Refs.~\cite{Papucci:2011wy,
lhcsusy,lhcsusy2,lhcsusy3,atlasSS,atlasSS2,atlastt,cmsSS,cmstt}). 
Vector (scalar) $X$ and ${\cal Z}$ particles appear in models with at least one (at least two)
universal extra dimension(s) \cite{ued}, stemming from the KK excitations of 
the multidimensional gauge-boson fields.\footnote{They can also occur in models with extended gauge groups \cite{Chivukula:1996yr}.}
Spin-0 color singlets and octets are
also found in ${\cal N}=2$ supersymmetry (SUSY) \cite{Polonsky:2000zt,n2,n2s}. 
Fermionic and vector top partners, $Y$, exist in extra-dimensional 
models \cite{ued} and SUSY models with an extended gauge 
sector \cite{Cai:2008ss}.
However, instead of focusing on specific particles in a particular model, we
pursue a general categorization in this paper, assuming only a discrete symmetry
that ensures the stability of $X$. 

Color octets with ${\cal O}$(TeV) masses can be pair produced with large cross
sections at the LHC. Consequently, within the framework of the MSSM, the ATLAS
and CMS collaborations have put strong bounds on their parameter space
\cite{atlasSS,atlasSS2,atlastt,cmsSS,cmstt}. In this paper, we recast these limits for
different spin assignments of the new particles. Despite these bounds, we
show that the full-energy run of the LHC (with 13--14 TeV collision energy)
will have a significantly expanded potential for searching for and possibly
discovering a signature of the type in Eq.~\eqref{eq:process}.
If a signal is observed, the next goal will be the determination of the spins
and couplings of the new particles, ${\cal Z}$, $Y$ and $X$.  We study
several observables for this purpose and demonstrate their usefulness
with numerical Monte Carlo simulation. To avoid
ambiguities due to model-dependent branching fractions, we do not rely on 
the total cross section in this set of variables. Looking further ahead, 
we also present the cross sections at the 100-TeV VLHC, where the signal 
production rates can be several orders of magnitude greater than the 14-TeV 
expectations and thus extend the discovery range substantially.

This paper is organized as follows. After introducing the model-independent
classification of gluon and top partners and discussing their production and
decay in section \ref{setup}, we show the current bounds and future reach of 
the LHC in sections \ref{current} and \ref{future}, respectively.
Section~\ref{prop} is devoted to the determination of the masses, spins and
couplings of the new particles from LHC data. Finally, conclusions are
presented in section~\ref{concl}.


\section{General Framework}
\label{setup}

\begin{figure}[tb]
\centering
\psfig{figure=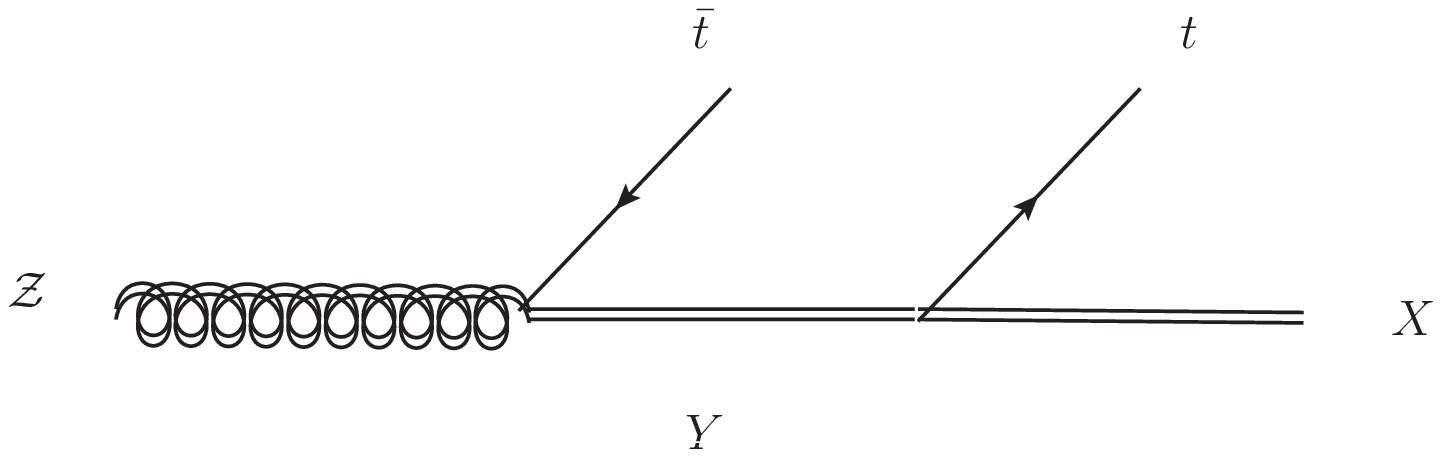, width=8.5cm}
\mycaption{The decay chain of ${\cal Z}$
to the color singlet $X$ via the color triplet $Y$. 
Double lines denote new particles, while single lines denote SM particles. 
If ${\cal Z}$ is a self-conjugate field, this decay chain is accompanied by the charge-conjugate version 
${\cal Z} \to t\bar{Y} \to t\bar{t}X$.
}
\label{fig:decay}
\end{figure}

In the following, we study the phenomenology of the Standard Model extended
by three new particles: a neutral
color singlet $X$, a color triplet $Y$ (and its antiparticle $\bar{Y}$) with
charge $+2/3$ ($-2/3$), 
and a color octet ${\cal Z}$. 
All new particles are assumed to be charged under some new global symmetry, 
so that they can be produced only in pairs and their decay chains end with 
the lightest new particle, which,  because of astrophysical limits, must be 
the $X$.
Searches for gluinos by ATLAS and CMS 
\cite{atlasSS,atlasSS2,cmsSS,atlastt,cmstt} have led to strong lower bounds 
on the mass of the ${\cal Z}$, so it is reasonable to assume the mass 
hierarchy $m_{\cal Z} > m_Y + m_t$, $m_Y > m_X + m_t$, leading to the decay 
chain shown in Fig.~\ref{fig:decay}.

\begin{figure}[tb]
\centering
\psfig{figure=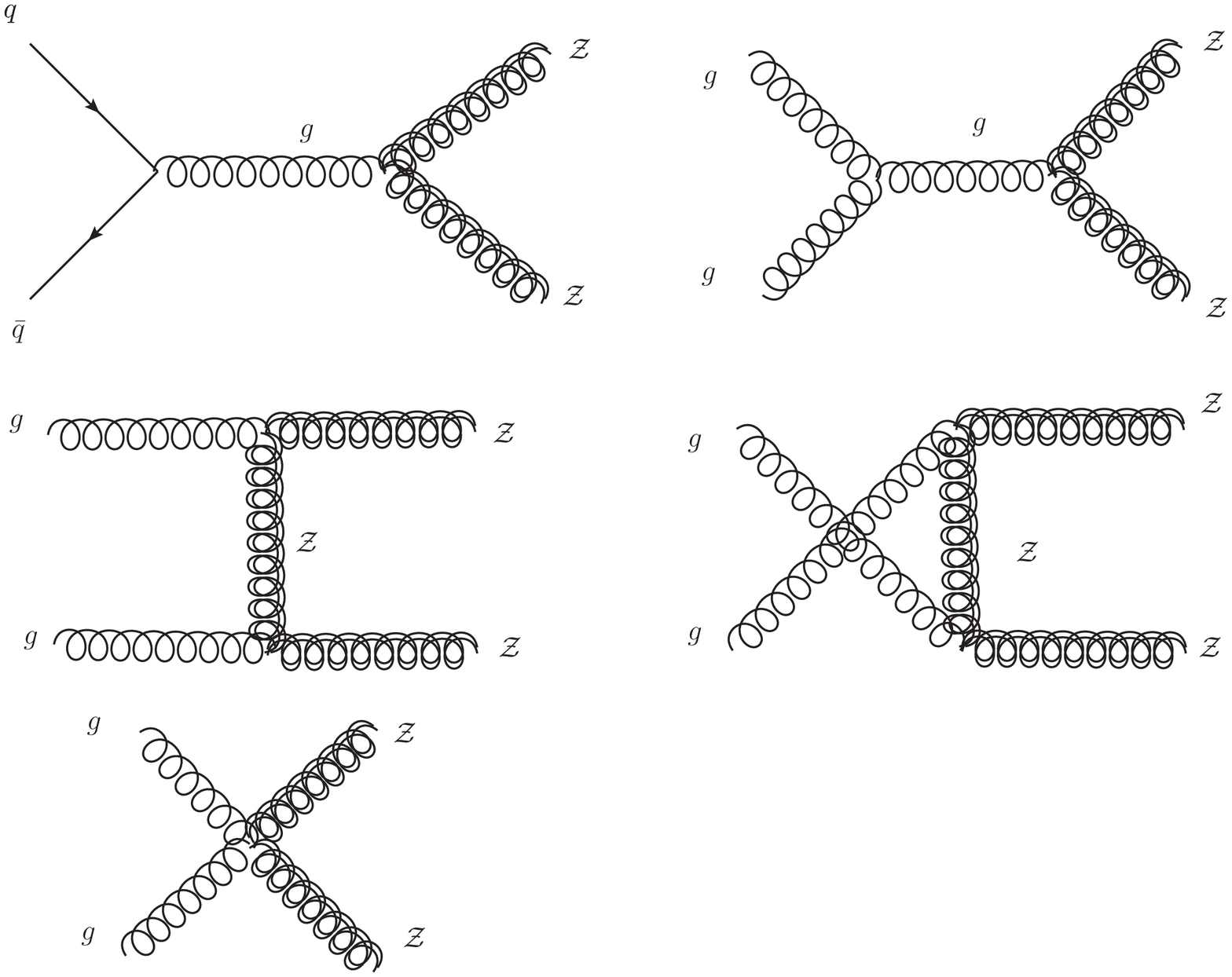, width=14.0cm}
\vspace{-.5em}
\mycaption{Leading-order diagrams for the pair production of the color octet
${\cal Z}$ at hadron colliders. Double lines denote new particles, while single lines 
denote SM particles. The last diagram exists only for a bosonic ${\cal Z}$.}
\label{fig:prodyz}
\end{figure}

When one demands gauge invariance and renormalizability, 
there are four possible spin combinations (with spin 0, 1/2 and 1) allowing a 
coupling between $Y$, $X$ and $t$, and four combinations for a coupling
between ${\cal Z}$, $Y$ and $t$. 
We summarize the possible combinations in Table~\ref{tab:models}. 
The top panel of the table (cases i$-$iv) reproduces the color-triplet 
interactions constructed in Ref.~\cite{cfhl}. These form the basis for our 
current extended theoretical framework including the color octet ${\cal Z}$ 
(cases v$-$viii).
By default, the $X$ and ${\cal Z}$ are assumed to be self-conjugate, but we 
also explore the phenomenological differences between the cases
where ${\cal Z}$ is a Majorana fermion (case vi(a)) and a Dirac fermion 
(case vi(b)), denoted ${\cal Z}_D$,
which can arise in models with ${\cal N}=2$ supersymmetry \cite{n2}. 
Also shown in the table are the structure of the relevant couplings and 
examples of concrete realizations of each case in specific models. For 
fermions, we allow a general chirality structure, specified by the 
parameters $a_{L,R}$ and $b_{L,R}$.

\begin{table}[tp]
\renewcommand{\arraystretch}{1.4}
\centering
\begin{tabular}{|l||c|c||c|c||ll|}
\hline
 & $Y$ & $X$ & $GYY$ & $XYt$ & 
 \multicolumn{2}{l|}{sample model and decay} \\[-1ex]
 & $s,\ I_{\rm SU(3)}$ & $s,\ I_{\rm SU(3)}$ &
 coupling & coupling & \multicolumn{2}{l|}{$Y \to t X$} \\
\hline\hline
i & 0, {\bf 3} & $\frac{1}{2}$, {\bf 1} & 
 $G^{a\mu} Y^* \!\overleftrightarrow{\partial}_{\!\!\!\mu} T^a Y $ &
 $\overline{X} \Gamma t \, Y^*$ &
 MSSM & $\tilde{t} \to t \tilde{\chi}_1^0$ \\
\hline
ii & $\frac{1}{2}$, {\bf 3} & 0, {\bf 1} & 
 $\overline{Y} \SLASH{G}{.3}\,^a T^a Y $ &
 $\overline{Y} \Gamma t \, X$ &
 UED & $t_{\rm KK} \to t \gamma_{H,\rm KK}$ \\
\hline
iii & $\frac{1}{2}$, {\bf 3} & 1, {\bf 1} & 
 $\overline{Y} \SLASH{G}{.3}\,^a T^a Y $ &
 $\overline{Y}\SLASH{X}{.4}\; \Gamma t$ &
 UED & $t_{\rm KK} \to t \gamma_{\rm KK}$ \\
\hline
iv & 1, {\bf 3} & $\frac{1}{2}$, {\bf 1} &
 $S_3[G,Y,Y^*]$ &
$\overline{X} \SLASH{Y}{.2}\,^* \Gamma t$ &
 \cite{Cai:2008ss} & $\vec{Q} \to t \tilde{\chi}^0_1$ \\
\hline
\multicolumn{7}{c}{}\\
\hline
 & ${\cal Z}$ & $Y$ & $G{\cal Z}{\cal Z}$ & ${\cal Z}Yt$ & 
 \multicolumn{2}{l|}{sample model and decay} \\[-1ex]
 & $s$, $I_{\rm SU(3)}$ & $s$, $I_{\rm SU(3)}$ & 
 coupling & coupling & \multicolumn{2}{l|}{${\cal Z} \to Y \overline{t}$} \\
\hline\hline
v & 0, {\bf 8} & $\frac{1}{2}$, {\bf 3} &
 $G^{a\mu} {\cal Z}^c \!\overleftrightarrow{\partial}_{\!\!\!\mu} {\cal Z}^b f^{abc}$ &
 $\overline{Y} T^a \Gamma' t {\cal Z}^a$ &
 UED & $g_H \to t_{\rm KK} \overline{t} $ \\
\hline
vi(a) & $\frac{1}{2}$, {\bf 8} & 0, {\bf 3} &
 $\overline{{\cal Z}}^c \SLASH{G}{.3}\,^a {\cal Z}^b f^{abc}$ &
$\overline{{\cal Z}}^a Y^* T^a \Gamma' t$ &
 MSSM & $\tilde{g} \to \tilde{t} \overline{t}$ \\
\hline
vi(b) & $\frac{1}{2}$, {\bf 8} & 0, {\bf 3} &
 $\overline{{\cal Z}}_D^c \SLASH{G}{.3}\,^a {\cal Z}_D^b f^{abc}$ &
\parbox[c][3em]{6.7em}{$(\overline{{\cal Z}}_D^a)^* Y^* T^a b_L t_L$\\[.5ex]
$+\;\overline{{\cal Z}}_D^a Y^* T^a b_R t_R$} &
\parbox[c]{1.5cm}{${\cal N}=2$\\ SUSY} & $\tilde{g}_D \to \tilde{t} \overline{t}$ \\
\hline
vii & $\frac{1}{2}$, {\bf 8} & 1, {\bf 3} &
 $\overline{{\cal Z}}^c \SLASH{G}{.3}\,^a {\cal Z}^b f^{abc}$ &
$\overline{{\cal Z}}^a \SLASH{Y}{.2}\,^* T^a \Gamma' t$ &
\cite{Cai:2008ss} & $\tilde{g} \to \vec{Q} \overline{t}$  \\
\hline
viii & 1, {\bf 8} & $\frac{1}{2}$, {\bf 3} &
 $S_8[G,{\cal Z},{\cal Z}]$ &
$\overline{Y} \SLASH{{\cal Z}}{0.4}^a T^a \Gamma' t$ &
 UED & $g_{\rm KK} \to t_{\rm KK} \overline{t}$ \\
\hline
\end{tabular}
\vspace{1ex}
\begin{tabular}{l}
{\small{$ 
\Gamma \equiv a_L P_L + a_R P_R \,,\qquad
\Gamma' \equiv b_L P_L + b_R P_R \,$ }}\\
{\small{$A \!\overleftrightarrow{\partial}_{\!\!\!\mu} B \equiv
A (\partial_\mu B) - (\partial_\mu A) B
$}}\\
{\small{$
S_3[G,Y,Y^*] \equiv 
  G_\mu^a \, Y^{*}_\nu \!\overleftrightarrow{\partial}^{\!\!\!\mu} T^{a} Y^{\nu} +
  G_\mu^a \, Y^{\mu*} \!\overleftarrow{\partial}^{\!\!\nu} T^{a} Y_{\nu} -
  G_\mu^a \, Y^{*}_\nu \!\overrightarrow{\partial}^{\!\!\nu} T^{a} Y^{\mu}
$}}\\
{\small{$
S_8[G,{\cal Z},{\cal Z}^*] \equiv f^{abc} \left[
  G_\mu^a \, {\cal Z}^{c*}_\nu \!\overleftrightarrow{\partial}^{\!\!\!\mu} {\cal Z}^{b\nu} +
  {\cal Z}_\mu^b \, G_\nu^a \!\overleftrightarrow{\partial}^{\!\!\!\mu} {\cal Z}^{c\nu*} +
  {\cal Z}^{c*}_\mu \, {\cal Z}_\nu^b \!\overleftrightarrow{\partial}^{\!\!\!\mu} G^{a\nu}
\right]
$
}}
\end{tabular}
\vspace{-1ex}
\mycaption{Quantum numbers and couplings of the new particles, $X$, $Y$ and
${\cal Z}$, which interact with the SM top quark, $t$. 
In the last column, $\tilde{g}$, $\tilde{t}$ and $\tilde{\chi}^0_1$ are the gluino, the scalar top and
lightest neutralino in the MSSM, respectively \cite{susy}.
$g_{\rm KK}$, $t_{\rm KK}$, 
$\gamma_{\rm KK}$, $g_{H}$ and $\gamma_{H,\rm KK}$ are the first-level Kaluza-Klein 
excitations of the gluon, the top, the photon, and an extra-dimensional component of a gluon
and a photon, respectively, in universal extra dimensions (UED) \cite{ued}.
$\tilde{g}_D$ denotes a Dirac gluino in ${\cal N}=2$ supersymmetry
\cite{n2,Choi:2008pi}.
Finally, $\vec{Q}$ is the vector superpartner in a supersymmetric model with 
an extended gauge sector \cite{Cai:2008ss}.}
\label{tab:models}
\end{table}


\vspace{\bigskipamount}

Direct production of $Y\bar{Y}$ pairs was discussed in detail in
Ref.~\cite{cfhl}. Here we consider pair production of ${\cal Z}$ particles, 
which subsequently decay according to the decay chain in Fig.~\ref{fig:decay}. 
They can be produced with sizeable cross sections at the LHC, even for large 
masses, $m_{\cal Z} > 1\tev$, and lead to a distinct final state of four top 
quarks and missing energy.
The dominant modes for ${\cal Z}$ pair production at hadron colliders are the 
QCD subprocesses
\begin{align}
 q \bar{q},\  gg & \rightarrow  {\cal Z} \bar{\cal Z} \,,
\end{align}
which are described at leading order by the diagrams
in Fig.~\ref{fig:prodyz}. The form of the gluon-${\cal Z}$ vertex is dictated 
by QCD gauge invariance and shown in
Table~\ref{tab:models}.
%
\begin{figure}
\anc\hspace{16mm}\makebox[8.5cm][l]{\small $\sqrt{s}=8$~TeV}%
\makebox[6.5cm][l]{\small $\sqrt{s}=14$~TeV}\\[.5ex]
\epsfig{figure=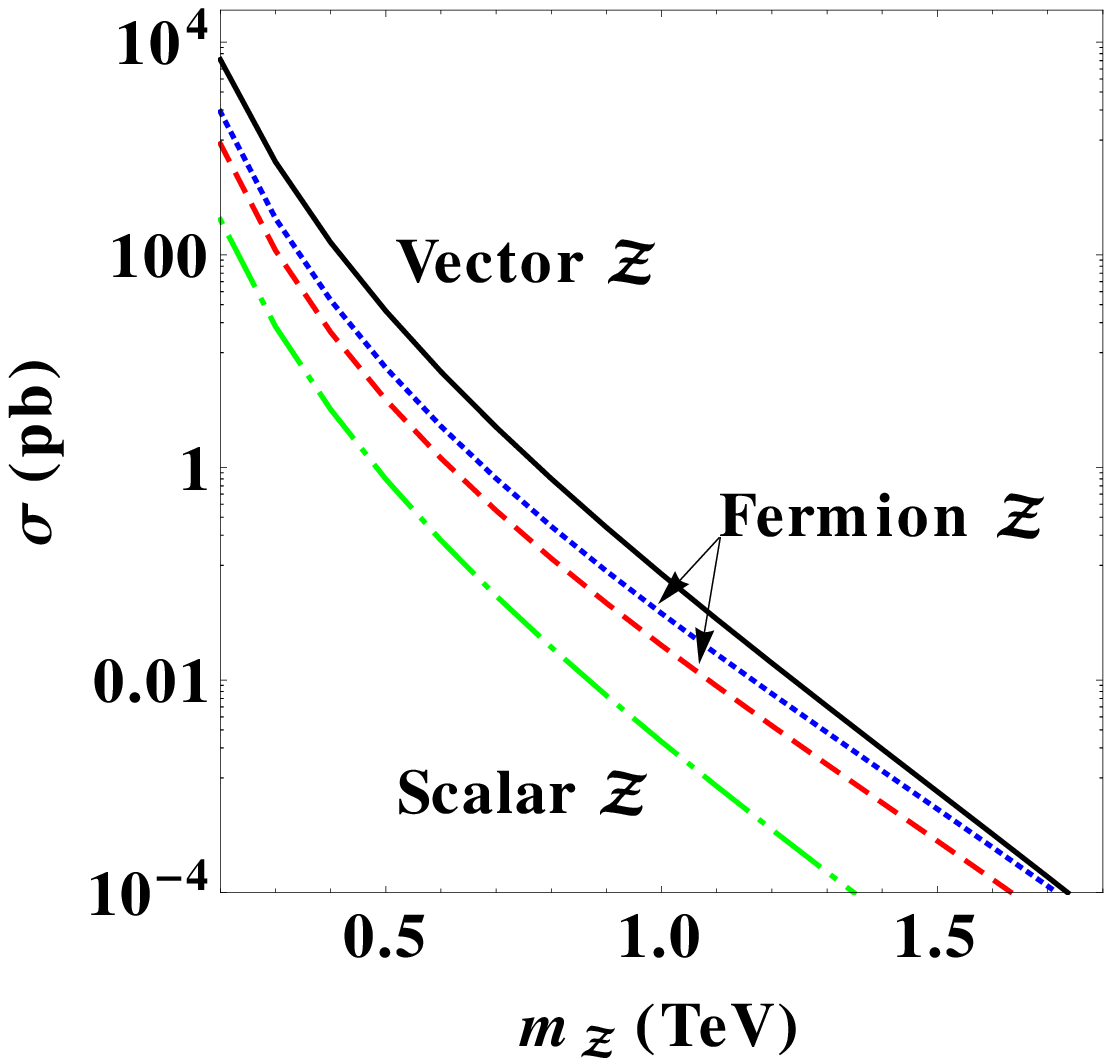, height=7.5cm}
\hfill
\epsfig{figure=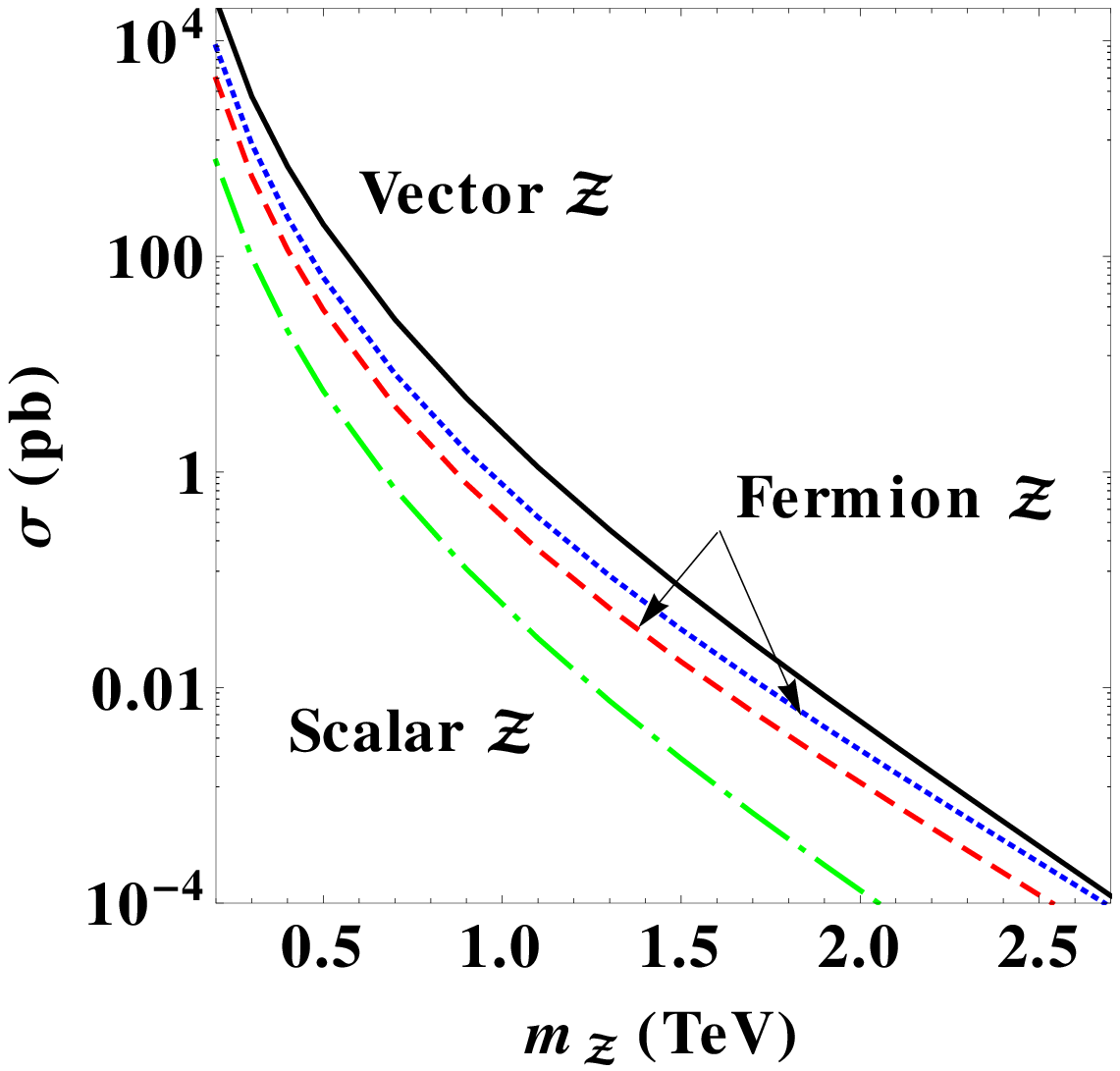, height=7.5cm}
\vspace{-.3em}
\mycaption{Production cross sections for $pp\to {\cal Z}\bar{{\cal Z}}$ at the 
LHC for
8~TeV (left) and 14~TeV (right), as a function of the mass $m_{\cal Z}$, 
for a vector ${\cal Z}$ (black solid), Dirac ${\cal Z}$ (blue dotted),
Majorana ${\cal Z}$ (red dashed), and scalar ${\cal Z}$ (green dot-dashed).
%
\label{plot:prodz}}
\end{figure}

For the different ${\cal Z}$ spin assignments,
the total QCD cross sections at the LHC are
shown in Fig.~\ref{plot:prodz} as a function of $m_{\cal Z}$.
The values include next-to-leading-order (NLO) QCD corrections for the scalar ${\cal Z}$ \cite{GoncalvesNetto:2012nt}, and NLO and
next-to-next-to-leading
logarithmic (NNLL) corrections for the Majorana fermion ${\cal Z}$ \cite{Langenfeld:2012ti},
with extrapolation to larger or smaller values of $m_{\cal Z}$
where necessary. For pair production of color-octet Dirac fermions and vectors,
the QCD corrections have not been calculated to the best of our knowledge.
Therefore, we simply assume that the $K$-factor for Dirac fermions is identical to the one for Majorana fermions, and that the $K$-factor for the spin-1 case is the same as for the scalar case, since scalars and vectors 
share the same diagrams.

As Fig.~\ref{plot:prodz} shows, the cross section for fermion ${\cal Z}$
pairs is about one order of magnitude larger than for the scalar case, because
of 
the fermion's larger number of spin degrees of freedom and $p$-wave suppression
of the scalar. The latter effect is most  pronounced near threshold,
where the $p$-wave production has a velocity dependence of $\sigma \sim \beta^3=(1-4m_{\cal Z}^{2}/\hat s)^{3/2}$, whereas an $s$-wave leads to $\sigma \sim \beta$. As a result, the
difference between the scalar and fermion cross sections increases at small
values of $\beta$, that is, for large values of $m_{\cal Z}$. The production
cross section for Dirac fermions is twice as large as for Majorana fermions,
since Dirac fermions have twice the number of independent degrees of freedom.
The production rate for a vector ${\cal Z}$ is larger than that for Majorana
fermions by another factor of about five.

\begin{figure}
\anc\hspace{5mm}\makebox[8.7cm][l]{(a)\hspace{3mm} {\small $\sqrt{s}=100$~TeV}}
\makebox[8cm][l]{(b)\hspace{3mm} {\small}}\\[-.6ex]
\epsfig{figure=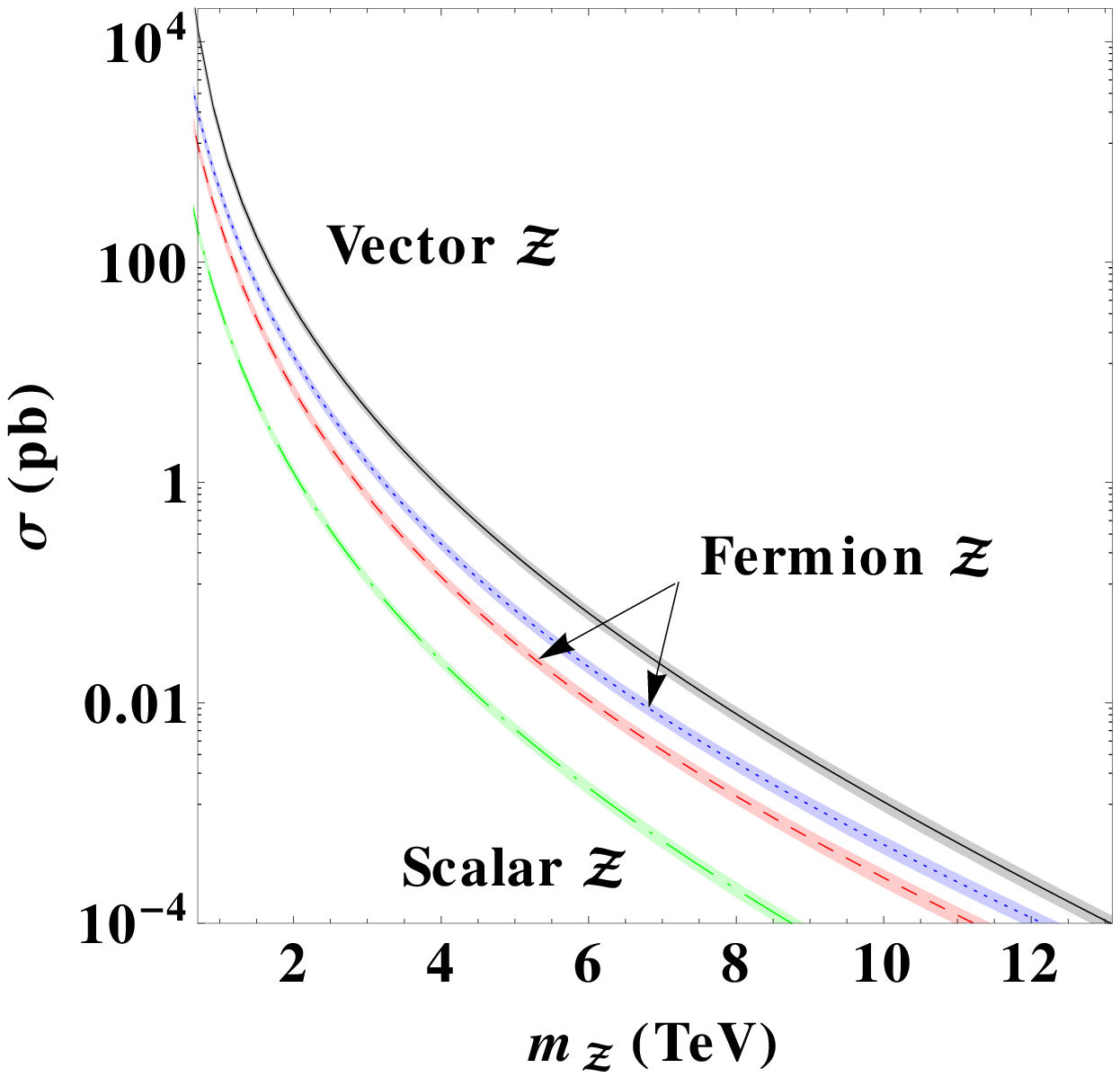, height=7.4cm}
\hfill
\epsfig{figure=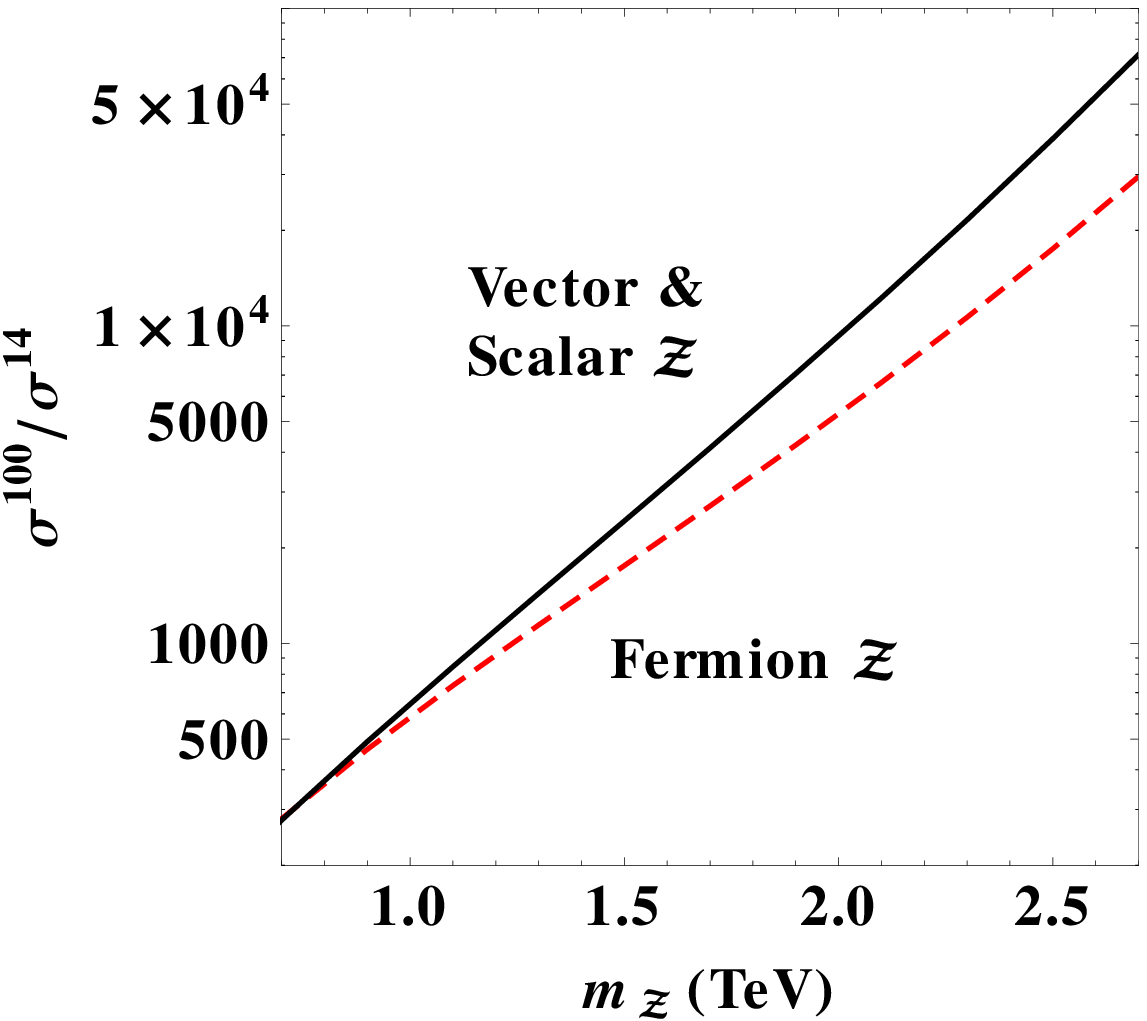, height=7.6cm}
\vspace{-.3em}
\mycaption{(a) Production cross sections for $pp\to {\cal Z}\bar{{\cal Z}}$
at a $pp$ collider with $\sqrt{s}=100$~TeV 
and (b) the cross-section ratios $\sigma(100\ {\rm TeV})/\sigma(14\ {\rm TeV})$, as a function of the mass $m_{\cal Z}$, for a vector ${\cal Z}$ (black solid), Dirac ${\cal Z}$ (blue dotted),
Majorana ${\cal Z}$ (red dashed), and scalar ${\cal Z}$ (green dot-dashed). The
widths of the bands indicate the estimated theoretical uncertainty
\cite{Borschensky:2014cia}.
\label{plot:vlhc}}
\end{figure}

For illustration, Fig.~\ref{plot:vlhc}(a) shows the total production
cross sections for $\cal Z\bar{Z}$ pairs at a proposed 100-TeV collider. 
We approximate the $K$-factors for scalar and vector $\cal Z\bar{Z}$ 
production at $\sqrt{s}=100$~TeV by assuming the same energy dependence 
as for the fermionic case, that is, we multiply their $K$-factors at 
$\sqrt{s}=14$~TeV by the ratio of the fermionic $K$-factors at 100~TeV
\cite{Borschensky:2014cia} and 14~TeV.
The shaded bands underlying each curve indicate the estimated theoretical 
uncertainty
due to parton distribution functions and the dependence on renormalization and
factorization scales, as estimated in Ref.~\cite{Borschensky:2014cia}.
For comparison with the LHC reach, we show the cross-section ratios at the two 
energies, $\sigma(100\ {\rm TeV})/\sigma(14\ {\rm TeV})$, in 
Fig.~\ref{plot:vlhc}(b). We see that the production cross sections for the 
color-octet particles could increase by a factor of 500$-$50,000 for a mass 
of 1$-$2.5~TeV. Thus, color octets with masses of ${\cal O}$(10 TeV) will 
become accessible at such a machine, which will reach a cross section of order 
0.1$-$1~fb. However, in the following sections, we shall focus
on the LHC phenomenology of these particles.

\section{Current Bounds from the 8-TeV LHC}
\label{current}

Processes of the form \eqref{eq:process} can be probed through LHC searches
for gluino production with the dominant decay $\tilde{g} \to t\bar{t}\tilde\chi^0_1$,
where $\tilde{\chi}^0_1$ is the lightest neutralino. In fact, the 
scenario vi(a)+i
in Table~\ref{tab:models} corresponds exactly to this MSSM process. For the other
cases in Table~\ref{tab:models}, one can obtain limits by recasting the
experimental MSSM results \cite{atlasSS,atlasSS2,atlastt,cmsSS,cmstt}. Some of
the strongest constraints are obtained from searches for multi-jet final states
\cite{atlastt,cmstt}. Here, we instead focus on searches with two same-sign
leptons in the final state \cite{atlasSS,atlasSS2,cmsSS}, which have a slightly
smaller mass reach but significantly less SM background. The reduced background is
an important advantage for model discrimination, which will be discussed in
section~\ref{prop}. In particular, we adopt the ATLAS analysis from
Ref.~\cite{atlasSS}, but the more recent paper \cite{atlasSS2} and the
CMS analysis in Ref.~\cite{cmsSS} lead to similar limits.

We have reproduced the simulation of the MSSM signal in Ref.~\cite{atlasSS}
using {\sc Pythia 6.4} \cite{pythia} and employing the selection cut sets SR1b
and SR3b from that analysis. Explicitly, these cuts are defined as follows:
\begin{align}
\text{Pre-sel.:} &&&\text{Two leptons with } p_{\rm T,\ell} > 20\gev, 
\; |\eta_e| < 2.47, \; |\eta_\mu| < 2.4, \text{ and same charge},
\label{eq:presel} \\
&&&N_j \text{ jets with } p_{{\rm T},j}>40\gev, \; |\eta_j| < 2.8,
\nonumber \\
&&&N_b \text{ $b$-jets with 70\% $b$-tagging efficiency and 1\%
light-jet mis-tagging rate}, \nonumber \\
&&&\Delta R_{\rm \ell\ell} > 0.3, \qquad \Delta R_{jj} > 0.4, \qquad
\Delta R_{\ell j} > 0.3. \nonumber \displaybreak[0]
\\[1ex]
\text{SR1b:} &&& N_j \geq 3, \quad N_b \geq 1, \label{eq:sr1b} \\
&&&|\SLASH{\mbox{\bf p}}{.2}_{\,T}|>150\gev, \quad M_{T}(\ell_1,\SLASH{\mbox{\bf p}}{.2}_{\,T}) > 100\gev,
\quad m_{\rm eff} > 700\gev. \nonumber
\\[1ex]
\text{SR3b:} &&&N_j \geq 4, \quad N_b \geq 3. \label{eq:sr3b}
\end{align}
Here, $p_{T}$ and $\eta$ denote the transverse momentum and pseudorapidity of
an object, respectively, $\Delta R_{ab} \equiv \sqrt{(\eta_a - \eta_b)^2 + (\phi_a - \phi_b)^2}$, and  
$\SLASH{\mbox{\bf p}}{.2}_{\,T}$ is the missing transverse momentum.  
The number $N_j$ includes both light jets and $b$-jets. The effective mass 
$m_{\rm eff} = \sum_\ell |\mbox{\bf p}_{ T,\ell}| + \sum_j |\mbox{\bf p}_{T,j}| + |\SLASH{\mbox{\bf p}}{.2}_{\,T}|$ 
is the scalar sum of the missing transverse momentum and
the transverse momenta of the selected leptons and jets, and
$M_{T}(\ell_1,\SLASH{\mbox{\bf p}}{.2}_{\,T}) = \sqrt{2|\mbox{\bf p}_{T,\ell_1}||\SLASH{\mbox{\bf p}}{.2}_{\,T}|
-2\mbox{\bf p}_{T,\ell_1}\cdot\SLASH{\mbox{\bf p}}{.2}_{\,T}}$ 
is the transverse mass associated with the leading lepton $\ell_1$. A cut on $M_{T}$
reduces the background from gauge-boson pair production.

After applying these cuts for $\sqrt{s}=8\tev$, we obtain event numbers that are
very similar to those in Table~5 of Ref.~\cite{atlasSS} (for an MSSM signal using the same
gluino, stop and neutralino masses as therein).

\begin{figure}[tb]
\centering
\psfig{figure=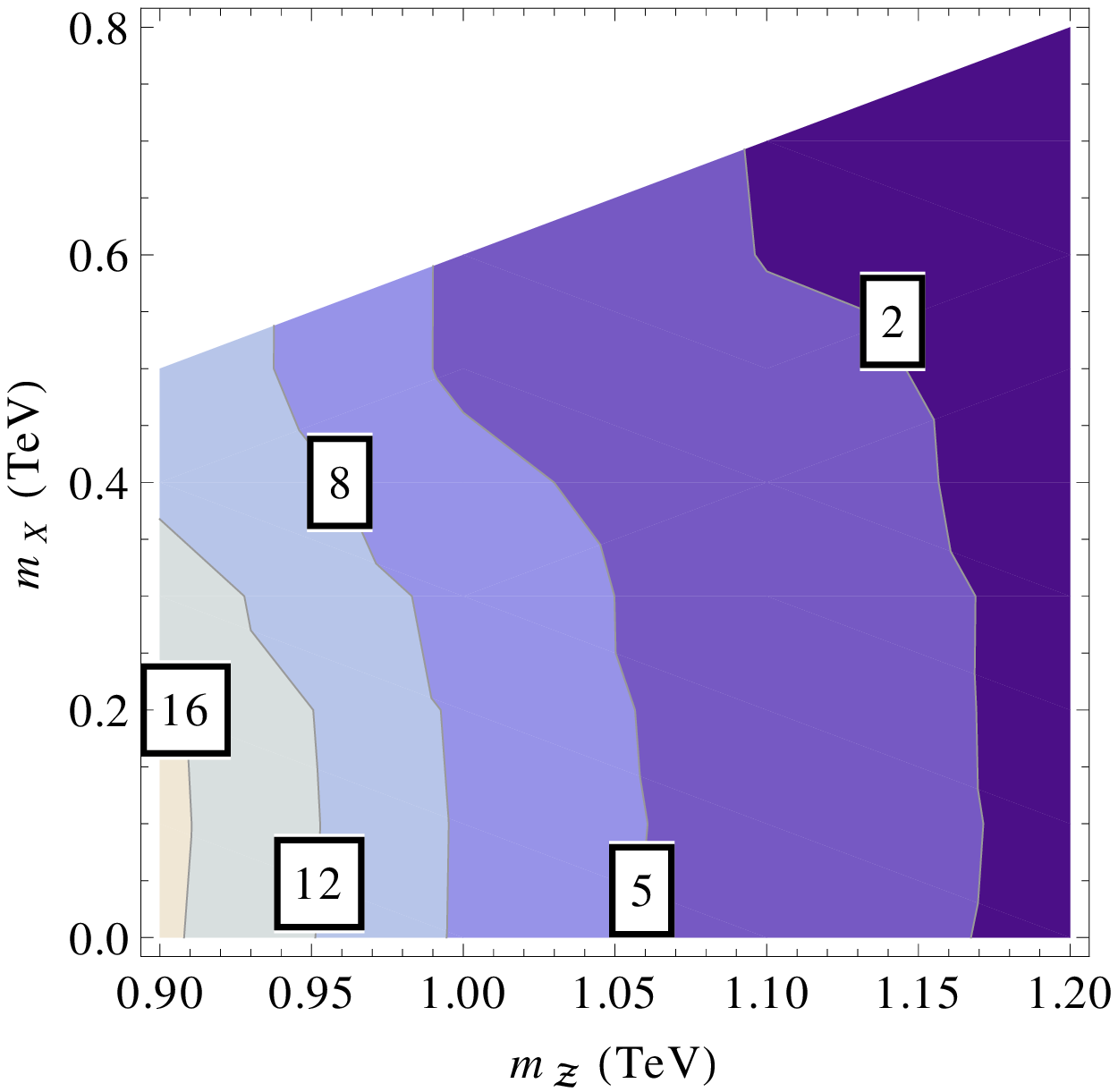, height=7.7cm}
\hfill
\psfig{figure=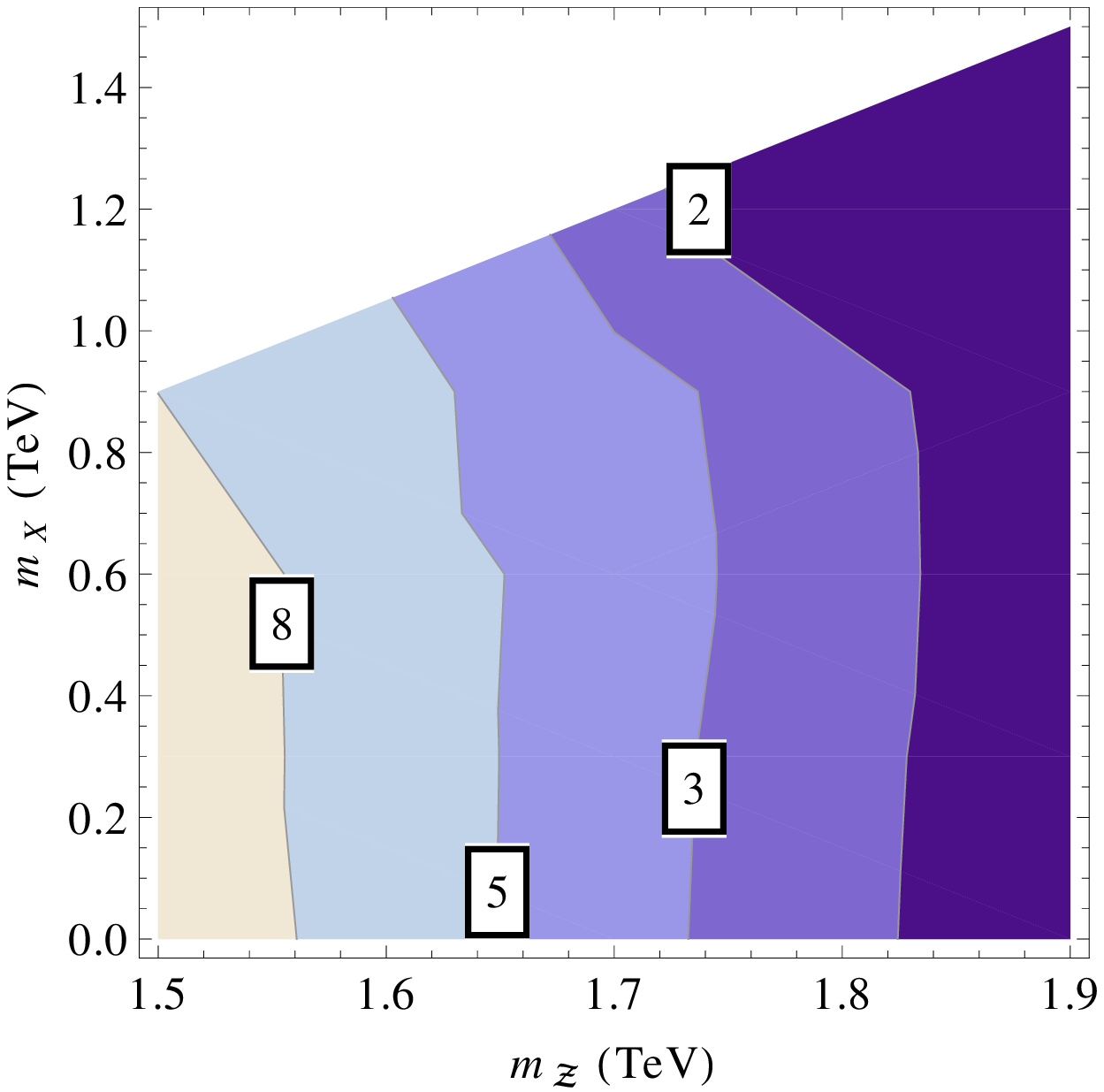, height=7.7cm}%
\vspace{-.5em}
\mycaption{Exclusion limits and projected discovery reach for $pp \to {\cal
Z\bar{Z}} \to t\bar{t}\,Y\bar{Y} \to t\bar{t}t\bar{t}X\bar{X}$ when ${\cal Z}$
is fermionic, as a function of the
masses of $\cal Z$ and $X$, with $m_Y=(m_{\cal Z}+m_X)/2$. 
The left panel corresponds to $\sqrt{s}=8\tev$ and
${\cal L}=21{\rm\ fb}^{-1}$, while the right panel corresponds to
$\sqrt{s}=14\tev$ and ${\cal L}=300{\rm\ fb}^{-1}$.
Contours are labeled with $\sigma$ values indicating the statistical significance.}
\label{fig:sf}
\end{figure}

The accurate evaluation of the SM backgrounds depends additionally on issues
like particle (mis)identification efficiencies; for these, we simply take the 
numbers from Table~3 in Ref.~\cite{atlasSS}. We then combine the SM
backgrounds with our simulation of the signal, for the case of the MSSM, which
corresponds to scenario vi(a)+i in Table~\ref{tab:models} (that is, the fermion-scalar-fermion 
spin combination), and perform a $\chi^2$ analysis. The results for
$\sqrt{s}=8\tev$ are shown in the left panel of Fig.~\ref{fig:sf}, as a function
of $m_{\cal Z}$ and $m_X$, with $m_Y=(m_{\cal Z}+m_X)/2$.
This sample value of $m_Y$ is representative of scenarios in which
neither of the ${\cal Z}$ and $Y$ decays is near threshold.

As Fig.~\ref{fig:sf} shows, fermionic octets (gluinos) decaying into top-quark
final states are excluded for $m_{\cal Z} \lesssim 1160\gev$.
This limit approximately agrees with Ref.~\cite{atlasSS}, although the detailed extent of
the excluded region depends on the choice of $m_Y$.


\section{Signal Observability at the 14-TeV LHC}
\label{future}

To obtain projections for $\sqrt{s}=14\tev$, we adjust the selection 
cuts in Eqs.~\eqref{eq:presel}--\eqref{eq:sr3b} to obtain roughly the
same signal efficiency as for $\sqrt{s}=8\tev$. Specifically, all cuts on 
dimensionless variables are left unchanged, while the cut
values for dimensionful variables are scaled up by a factor of 1.1
\footnote{Checking a range of points throughout the parameter space,
we find that the signal efficiencies agree to within 10\%, which is within 
the overall uncertainty of our analysis.}. 
We assume that, with this rescaled set of cuts, the same percentage of SM 
background events is retained as at $\sqrt{s}=8\tev$ with the original set 
of cuts, Eqs.~\eqref{eq:presel}--\eqref{eq:sr3b}. 
In other words, we estimate the SM background by scaling the event numbers 
from Ref.~\cite{atlasSS} by the ratio of the total cross sections for 
$\sqrt{s}=14\tev$ and $\sqrt{s}=8\tev$. The cross sections for the dominant 
SM processes, $pp\to t\bar{t}W,t\bar{t}Z$ and $pp \to WZ,ZZ$, are
taken from Refs.~\cite{Campbell:2012dh,Garzelli:2012bn,Campbell:2011bn}.

Using this procedure, we obtain the estimated reach of the 14-TeV LHC for the fermion-scalar-fermion 
spin combination given in the right panel of Fig.~\ref{fig:sf}.
Our results are consistent with Fig.~52 in Ref.~\cite{lhcsusy3}, although in 
that reference a different set of cuts has been used and the scalar $Y$ 
(stop) has been decoupled (that is, $m_Y \to \infty$).

\begin{figure}[tbp]
\centering
\psfig{figure=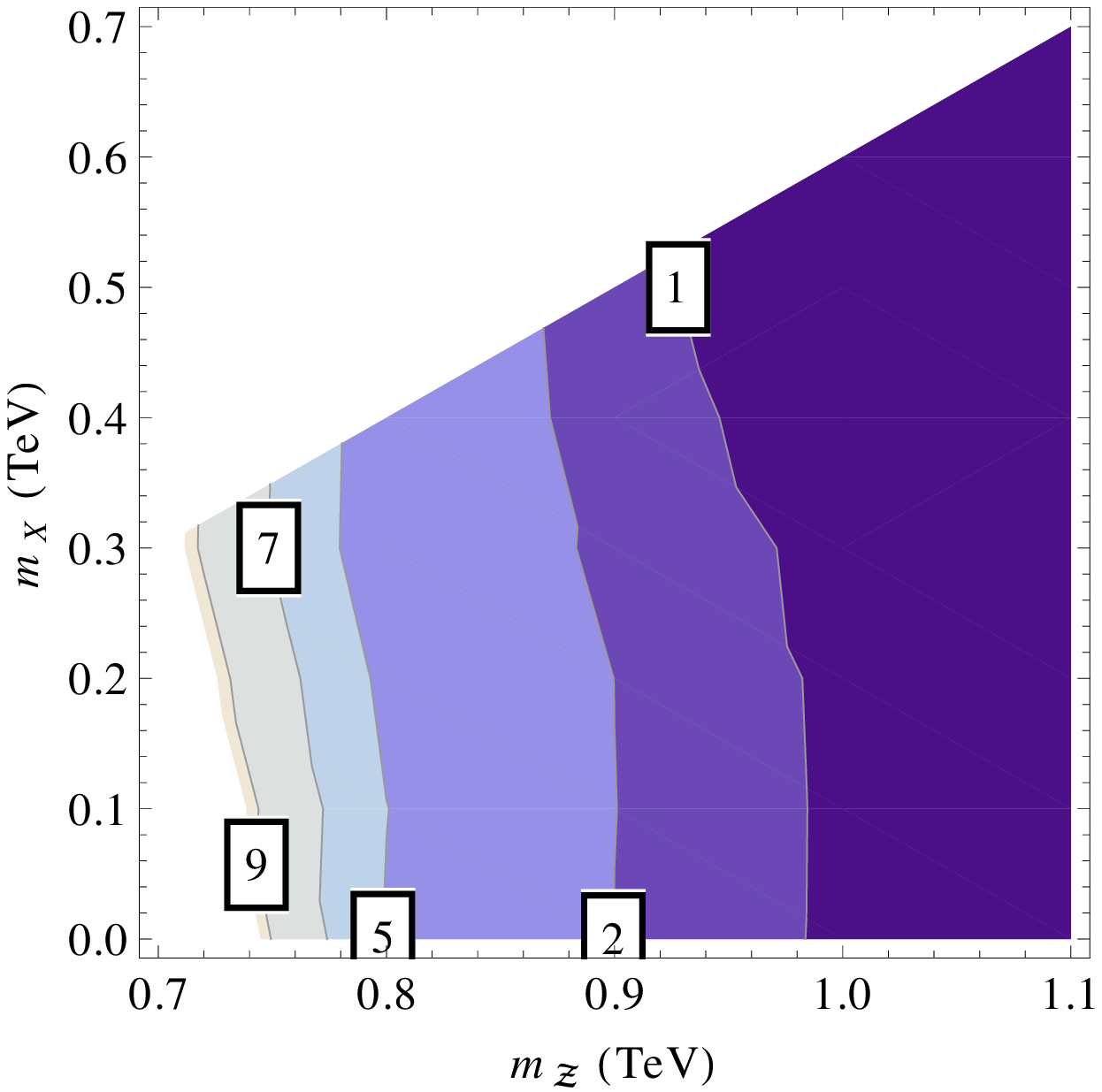, height=7.7cm}
\hfill
\psfig{figure=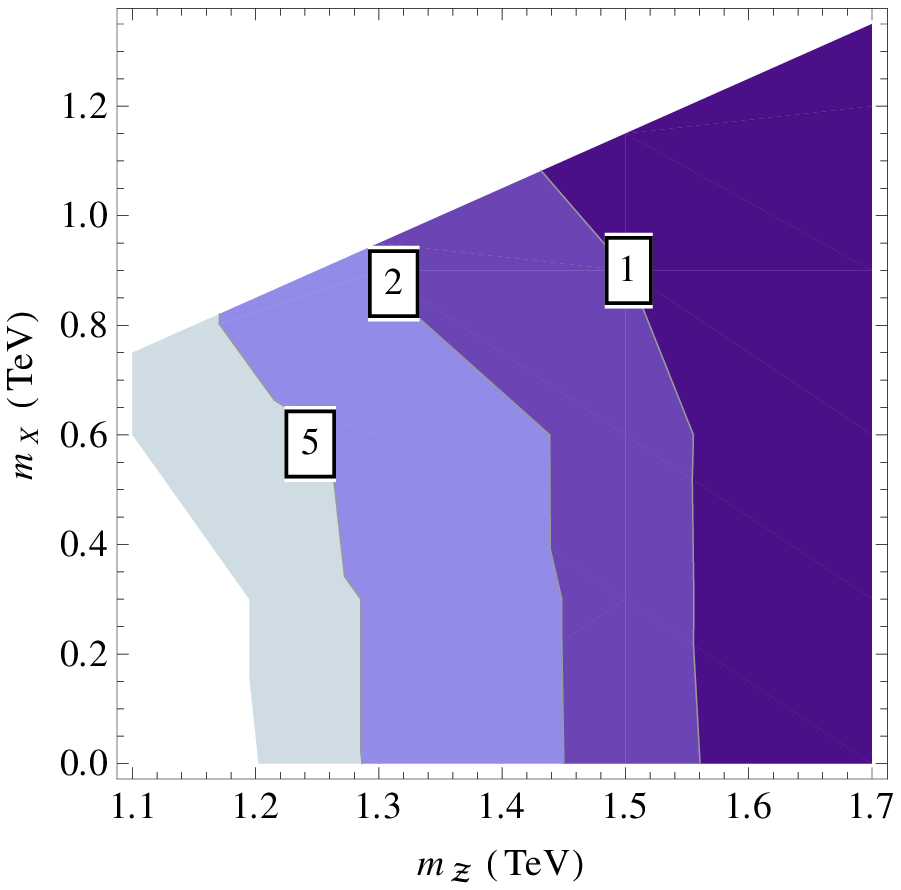, height=7.7cm}%
\\[1ex]
\psfig{figure=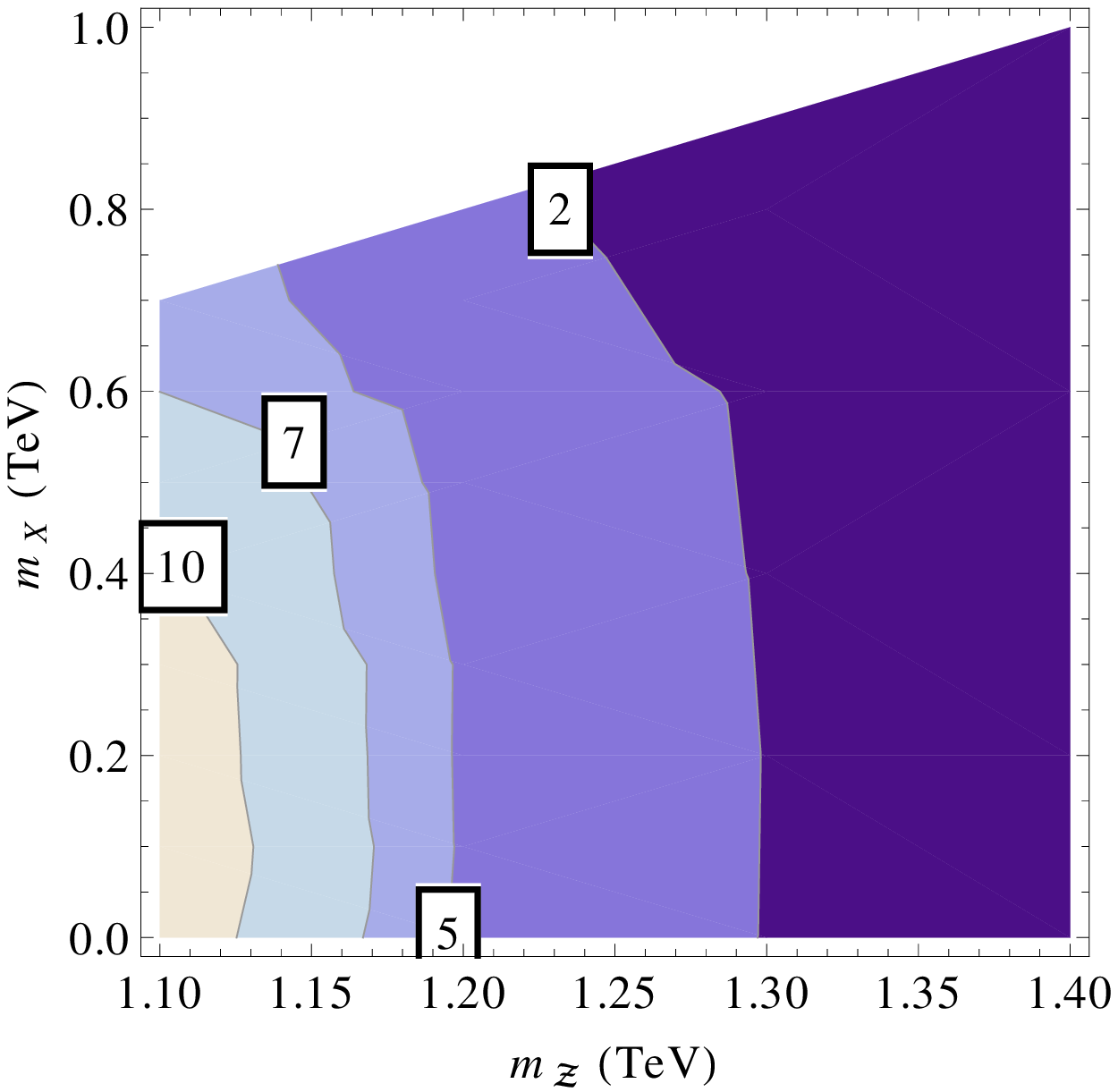, height=7.7cm}
\hfill
\psfig{figure=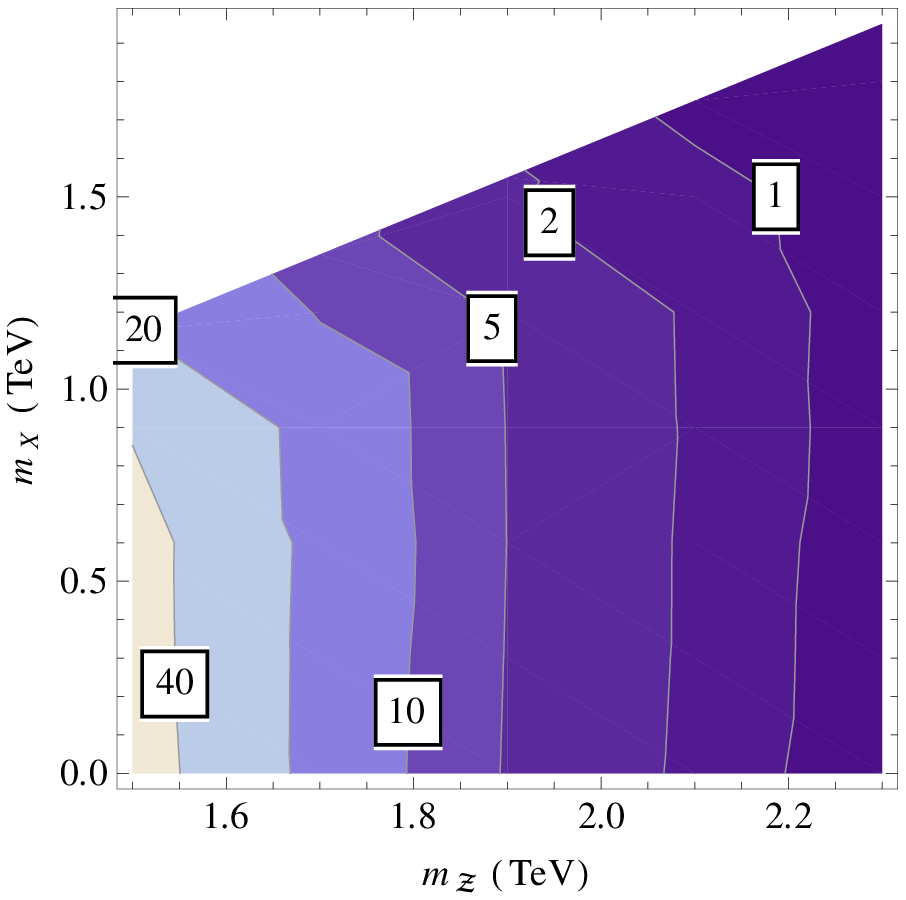, height=7.7cm}%
\vspace{-.5em}
\mycaption{Exclusion limits and projected discovery reach for $pp \to {\cal
Z\bar{Z}} \to t\bar{t}Y\bar{Y} \to t\bar{t}t\bar{t}X\bar{X}$ for a scalar 
(upper) and a vector (lower) ${\cal Z}$, as a function of the
masses of $\cal Z$ and $X$, with $m_Y=(m_{\cal Z}+m_X)/2$. 
The left panels correspond to $\sqrt{s}=8\tev$ and
${\cal L}=21{\rm\ fb}^{-1}$, while the right panels correspond to
$\sqrt{s}=14\tev$ and ${\cal L}=300{\rm\ fb}^{-1}$.
Contours are labeled with $\sigma$ values indicating the statistical 
significance.}
\label{fig:ssv}
\end{figure}

\vspace{\medskipamount}
The exclusion limits (for existing $\sqrt{s}=8\tev$ data) and projected reach
(for $\sqrt{s}=14\tev$) depend strongly on the spin of the ${\cal Z}$, because 
of its impact on the cross section $\sigma(pp \to {\cal
ZZ})$. One can obtain approximate limits for scalar and vector ${\cal Z}$ 
particles by rescaling the results in Fig.~\ref{fig:sf} by the relevant 
ratios of the cross sections shown in Fig.~\ref{plot:prodz}. Here, it is 
assumed that spin correlations in the decay chain 
${\cal Z} \to t\bar{Y}/\bar{t}Y \to t\bar{t}X$ 
have a small effect on the experimental selection efficiency, so that they 
can be neglected. 
The results are shown in Fig.~\ref{fig:ssv}. Note that one can obtain limits 
for the high-luminosity LHC with $\sqrt{s}=14\tev$ and 
${\cal L}=3000{\rm\ fb}^{-1}$ from the right panels in the figure by rescaling 
the contours by a factor of $\sqrt{10}$, under the assumption that statistical 
errors remain dominant.

From Fig.~\ref{fig:ssv}, one can extract the approximate $2\sigma$ exclusion 
limits for scalar and vector ${\cal Z}$ production at 8~TeV. For a 
light dark-matter candidate ($m_X \lesssim 200\gev$), the bounds are shown in 
Table~\ref{tab:reach}. The table also lists the expected reach of the 14-TeV 
run of the LHC for observation of the signal process in Eq.~\eqref{eq:process}  at the $5\sigma$ level, again assuming a light dark-matter candidate 
($m_X \lesssim 300\gev$).

\begin{table}[t]
\renewcommand{\arraystretch}{1.3}
\centering
\begin{tabular}{|c|c|c|c|}
\cline{2-4}
   \multicolumn{1}{c|}{}
 & spin-0 ${\cal Z}$ & spin-1/2 & spin-1 \\
\hline
 8 TeV (2$\sigma$ with 21 fb$^{-1}$)  & $  900 \gev$ &   $ 1160 \gev$  & $ 1290 \gev$ \\
 14 TeV  (5$\sigma$ with 300 fb$^{-1}$) & $ 1280 \gev$  & $ 1650 \gev$  & $ 1900 \gev$  \\
 14 TeV  (5$\sigma$ with 3000 fb$^{-1}$) & $ 1480 \gev$  & $ 1860 \gev$  & $ 2100 \gev$  \\
\hline
\end{tabular}
\mycaption{The 2$\sigma$ exclusion limit at 8~TeV and 5$\sigma$ discovery reach at 14 TeV 
for a spin-0, spin-1/2 and spin-1 ${\cal Z}$, assuming $m_X \lesssim 200\ (300)\gev$ for $\sqrt{s}=8$ $(14)$ TeV.}
\label{tab:reach}
\end{table}


\section{Determination of Model Properties}
\label{prop}

Once a new-physics signal consistent with the process in Eq.~\eqref{eq:process} has been observed at the LHC, it will be crucial to determine the particle properties in order to uncover the underlying theory. The kinematical distributions of the
final-state particles can be used to determine the masses, spins and couplings of
the $X$, $Y$ and $\cal Z$ particles. The analysis of direct pair
production of the color triplet $Y$, $pp \to Y\bar{Y} \to t\bar{t}X\bar{X}$, can
already yield valuable information about the properties of $Y$ and the singlet
$X$ \cite{cfhl}. In this section, we shall instead be concerned primarily with
the determination of the properties of the color octet $\cal Z$ from the
process \eqref{eq:process}. 

As in the previous sections, we shall focus on
the same-sign lepton signature, where each of the directly produced
color octets decays through one leptonically and one hadronically decaying top
quark, ${\cal Z\bar{Z}} \to t_\ell t_\ell \bar{t}_h \bar{t}_h  + \Eslash$ or
${\cal Z\bar{Z}} \to \bar{t}_\ell \bar{t}_\ell t_h t_h  + \Eslash$
(where $\Eslash$ denotes missing transverse energy).
Since this signal has small SM backgrounds, we shall neglect them in the
following, in order to highlight more clearly the differences between the
various scenarios in Table~\ref{tab:models}. Of course, in a detailed
experimental or phenomenological analysis, the SM background contamination and
its uncertainty will need to be accounted for, but we leave this for future
work.

\subsection{Masses}

The distribution of the invariant mass, $m_{t\bar{t}}$, of the top-antitop pair
from the decay chain ${\cal Z} \to t \bar{t} X$ (see Fig.~\ref{fig:decay}) has a
sharp endpoint at
\begin{equation}
(m_{t\bar{t}}^{\rm max})^2 = \frac{(m_{\cal Z}^2 - m_Y^2)(m_Y^2 - m_X^2)}{m_Y^2}.
\label{eq:endpoint}
\end{equation}
Even if one of the top quarks decays leptonically, the invariant-mass
distribution of the visible $t\bar{t}$ decay products ($b\bar{b}jj\ell$) still
has the endpoint in Eq.~\eqref{eq:endpoint}, but with a shallower
slope. In addition to measuring $m_{t\bar{t}}^{\rm max}$, one could
obtain information about $m_X$ and $m_Y$ from the process $pp \to Y\bar{Y} \to
t\bar{t}X\bar{X}$, using the observable $M_{\rm T2}$ or one of its variants
\cite{cfhl,mass2,mass}. By combining these observables, one could in principle
determine $m_X$, $m_Y$ and $m_{\cal Z}$ independently, albeit with poor
precision.

If instead one focuses on the all-hadronic decay channel of the top quark, so
that all top momenta can be reconstructed, one can take advantage of the
kinematical method in Ref.~\cite{Cheng:2007xv}, 
which gives relatively large errors in $m_X$ 
but fairly good precision for $m_{\cal Z}$ and $m_Y$. However, the separation of the
four top quarks in a given event is a difficult problem, which may be aided by
the use of a top-tagging algorithm (see, for example, Ref.~\cite{toptag}). 
Firmer conclusions will require a detailed simulation, which is beyond 
the scope of this paper.

\subsection{Spin}

\begin{figure}[tb!]
\psfig{figure=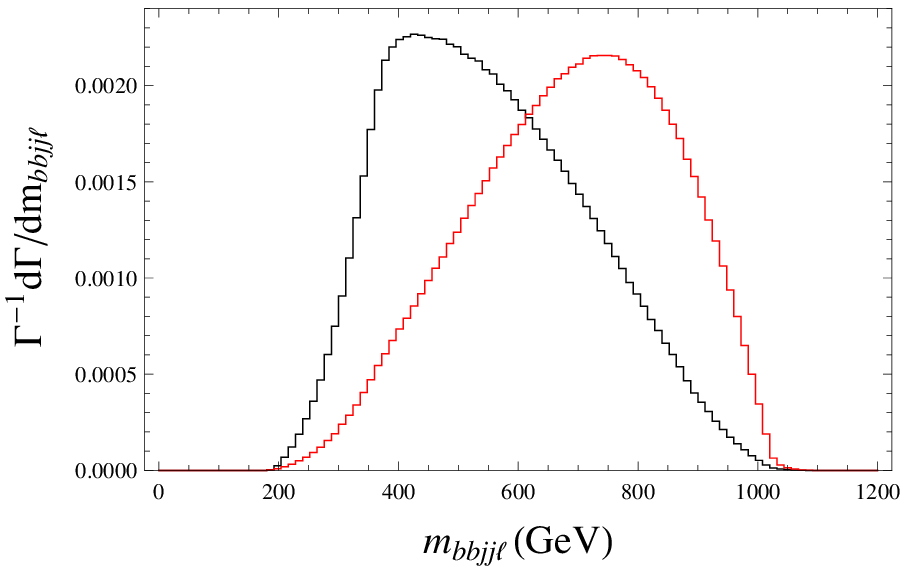, width=8.5cm, bb=0 25 260 162, clip=true}%
\rput[l](-6.9,4.1){
v+ii}\rput[l](-7,3.6){(SFS)}%
\psfig{figure=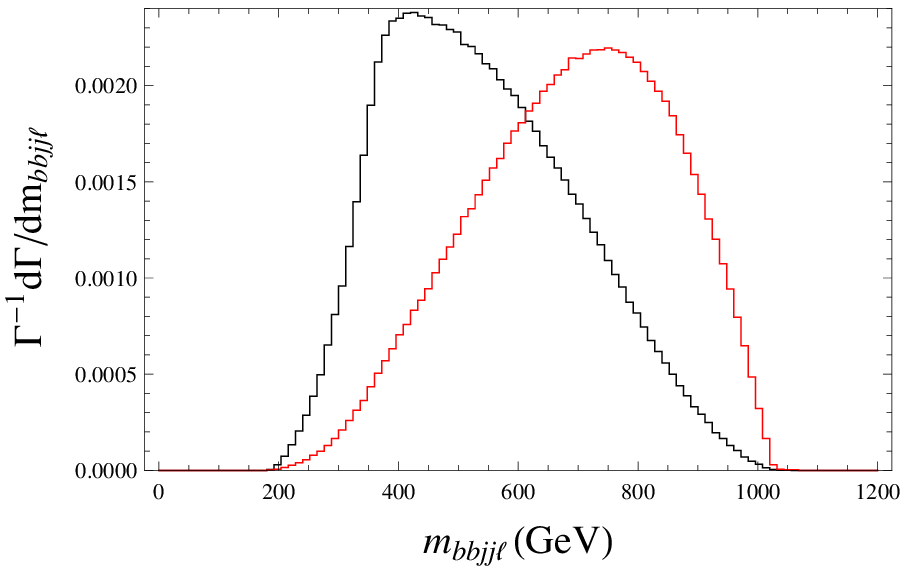, width=7.9cm, bb=18 25 260 162, clip=true}%
\rput[l](-6.8,4.1){v+iii}\rput[l](-6.9,3.6){(SFV)}\\
\psfig{figure=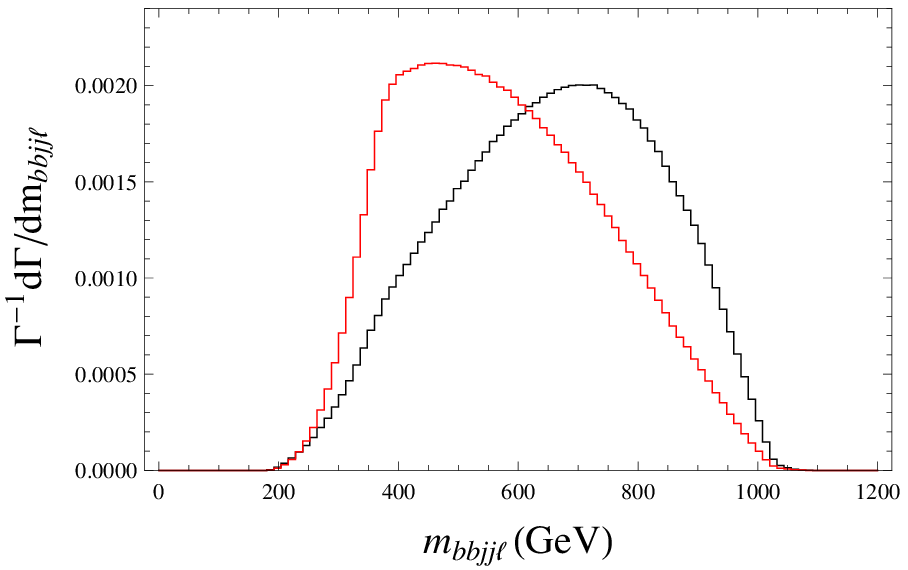, width=8.5cm, bb=0 25 260 162, clip=true}%
\rput[l](-6.9,4.1){viii+ii}\rput[l](-6.9,3.6){(VFS)}%
\psfig{figure=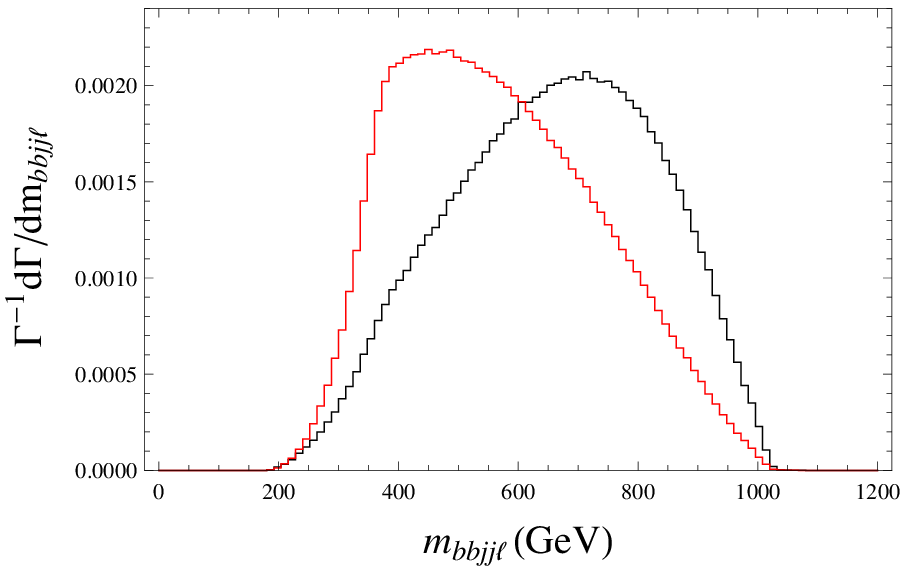, width=7.9cm, bb=18 25 260 162, clip=true}%
\rput[l](-6.9,4.1){viii+iii}\rput[l](-6.9,3.6){(VFV)}\\
\psfig{figure=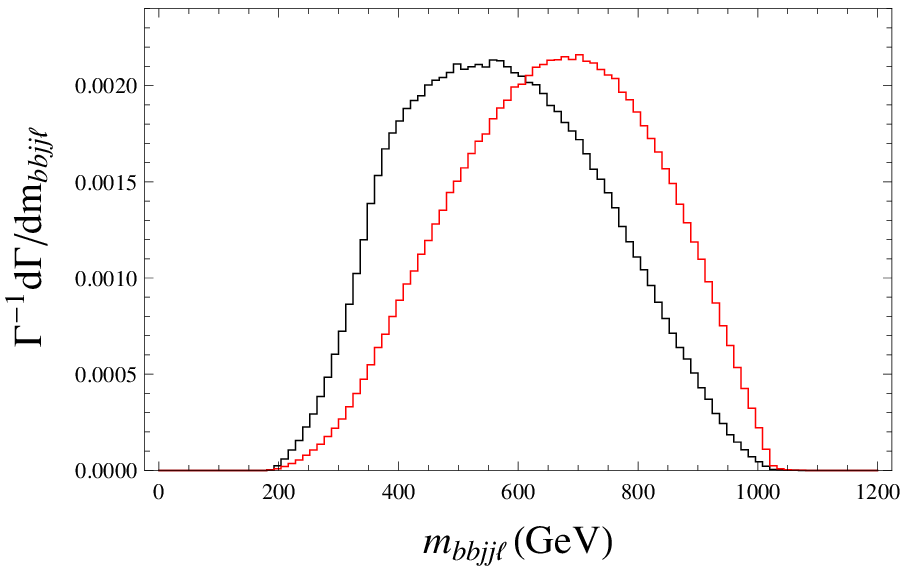, width=8.5cm}%
\rput[l](-6.9,4.9){vii+iv}\rput[l](-6.9,4.4){(FVF)}%
\psfig{figure=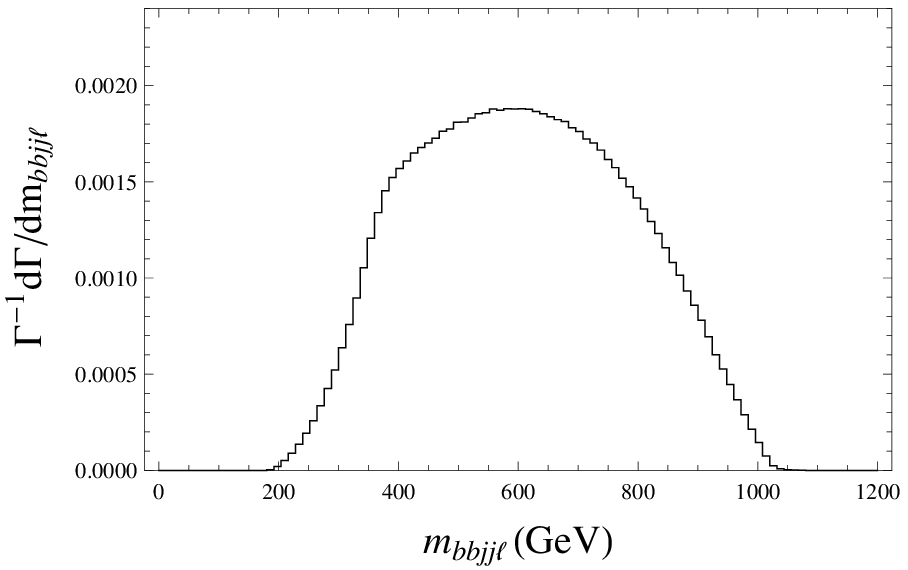, width=7.9cm, bb=19 0 260 162, clip=true}%
\rput[l](-6.9,4.9){vi(a)+i (FSF)}%
\vspace{-.5em}
\mycaption{Parton-level invariant-mass distribution of the visible decay 
products, for the decay chain ${\cal Z} \to t\bar{Y}/\bar{t}Y \to t\bar{t}X$.
The six panels show the results for the scenarios in Table~\ref{tab:models}, 
for the two coupling choices $a_L=1,\,a_R=0,\,b_L=1,\,b_R=0$ (black) 
and $a_L=0,\,a_R=1,b_L=1,\,b_R=0$ (red). Here, S, F, and V denote scalar, 
fermion, and vector particles, respectively, in the decay chain. 
The input mass parameters are 
$m_{\cal Z}=1200\gev$, $m_Y=600\gev$ and $m_X=100\gev$. 
The distributions have been normalized to unity.
\label{fig:mtt}}
\end{figure}

For a decay chain of the form in Fig.~\ref{fig:decay}, one can obtain
information about the spins of the ${\cal Z}$, $Y$, and $X$ particles from 
spin-correlation effects, which are reflected in the shape of the $t\bar{t}$
invariant-mass distribution. This strategy has been studied extensively
for similar decay chains involving leptons instead of top quarks
\cite{ll,spin1,Burns:2008cp}. In contrast to these studies, one must account
for the non-negligible mass of the top quark. Secondly, in focusing on the
same-sign lepton signature, one cannot fully reconstruct the $t\bar{t}$ mass 
because of the missing neutrino from the leptonic top decay.  Instead,
one has to work with the visible decay products of each decay chain, that is,
two $b$ jets, two light jets, and one charged lepton $\ell=e,\mu$. The invariant
mass of these objects, $m_{bbjj\ell}$, will have a distribution with similar
qualitative features to the $m_{t\bar{t}}$ distribution and thus can be used
for spin discrimination. 
Implementing the different spin combinations in Table~\ref{tab:models}
in {\sc CalcHEP} \cite{calchep} model files, we have performed 
parton-level simulations of the decay chain of a single $\cal Z$ particle,
obtaining Fig.~\ref{fig:mtt}.

For the fermion-scalar-fermion chain (scenarios vi$-$i), there are no spin 
correlations between the first and second step of the decay chain in 
Fig.~\ref{fig:decay}, so the $m_{t\bar{t}}$ distribution follows the shape 
dictated by the pure phase-space kinematics. As a result, for this case, 
$d\Gamma/dm_{bbjj\ell}$ 
peaks at medium values of $m_{bbjj\ell}$. 

In contrast, the scalar-fermion-scalar chain (scenario v$-$ii) displays maximal 
correlation effects in the $m_{t\bar{t}}$ distribution, 
since in this case angular momentum
conservation demands alignment between the $t$ and $\bar{t}$ helicities.
If the top and anti-top are produced with the same helicity
(corresponding to the choices $a_L=b_R=1, \; a_R=b_L=0$ or $a_L=b_R=0, \;
a_R=b_L=1$), then they are emitted preferentially in opposite directions, 
so that their spins add up to zero total angular momentum, as is necessary for 
the spin-0 initial ${\cal Z}$. 
As a result, in this case the invariant-mass distribution peaks at
large values of $m_{bbjj\ell}$. On the other hand, for opposite helicities of
the top and anti-top (that is, $a_L=b_L=1, \; a_R=b_R=0$ or $a_L=b_L=0, \;
a_R=b_R=1$), they are emitted mostly in the same direction, and thus the
$m_{bbjj\ell}$ distribution peaks at small values.

If $m_X/m_Y \ll 1$, the results for the scalar-fermion-vector chain  (scenario v$-$iii)
are very similar, since the excitation of different spin states of the vector $X$
is suppressed by $m_X/m_Y$ \cite{spin1}.
On the other hand, for scenario viii$-$ii or viii$-$iii (vector-fermion-scalar 
or vector-fermion-vector), the spin-correlation effects in the $m_{t\bar{t}}$
distribution are slightly reduced, since angular momentum conservation now
involves the helicity states of not only the $t$ and $\bar{t}$ but also the
parent ${\cal Z}$. The correlation effects are even further washed out for scenario
vii (fermion-vector-fermion), where both the initial ${\cal Z}$ and final $X$
have non-trivial spin states.

To study quantitatively how well one can distinguish between the different spin
combinations, we have performed a $\chi^2$ analysis for the binned $m_{bbjj\ell}$
distributions, using three bins\footnote{Larger numbers of bins do not yield
additional information, but only reduce the discriminative power because of the
increased number of degrees of freedom in the statistics.}. The resulting
$\sqrt{\chi^2}$ values are shown in Table~\ref{tab:spinchisq} for an assumed 
signal sample of $857$ events. This event yield corresponds to production of 
Majorana fermion pairs $\cal Z\bar{Z}$ with $m_{\cal Z}=1200$~GeV at $\sqrt{s}=14\tev$ with an
integrated luminosity of $300 \, \textrm{fb}^{-1}$. The number of events has
been obtained from the simulation results of section~\ref{future} and the total
cross section in Ref.~\cite{Langenfeld:2012ti}. 
\begin{table}[t!]
\renewcommand{\arraystretch}{1.25}
\centering
\newcolumntype{C}{>{\centering\arraybackslash}X}
\begin{tabularx}{.85\textwidth}{|c|CCCCC|}
\cline{2-6}
   \multicolumn{1}{c|}{}
 & \multicolumn{5}{c|}{Spin combinations}
  \\[-3pt]
   \multicolumn{1}{c|}{}
 & SFV & VFS & VFV & FVF & FSF \\
\hline
SFS & 1.3 & 10.3 & 10.6 & 2.4 & 5.3 \\[2pt]
SFV &  & 11.4 & 11.8 & 3.5 & 6.4 \\[2pt]
VFS &  &  & 0.35 & 9.6 & 4.7 \\[2pt]
VFV &  &  &  & 9.6 & 4.8 \\[2pt]
FVF &  &  &  &  & 3.6 \\[2pt]
\hline
\end{tabularx}
\mycaption{$\sqrt{\chi^2}$ values for the discrimination between pairs of different 
spin combinations, from a binned analysis of the invariant-mass distribution of
the visible $t\bar{t}$ decay products. Here, S, F, and V denote scalar, fermion, and vector
particles, respectively, in the decay chain. The results are based on 857 events for
the following input mass and coupling parameters: 
$m_{\cal Z}=1200\gev$, $m_Y=600\gev$, $m_X=100\gev$; 
$a_L=1,\,a_R=0,\,b_L=1,\,b_R=0$.}
\label{tab:spinchisq}
\end{table}

As Table~\ref{tab:spinchisq} shows, most pairs of spin combinations can be 
discriminated with high significance. An exception is pairs that differ only 
in the spin of the invisible $X$. Note that this analysis 
does not account for detector smearing effects, SM backgrounds, and 
combinatorial ambiguities in assigning the visible object in a given event 
to the decay chains of the $\cal Z$ and $\cal \bar{Z}$.


\subsection{Couplings}

The observable invariant-mass
distribution depends not only on the spin of the particles in the decay chain
but also on the chiral structure of their couplings, that is, whether they are
left- or right-handed (see Fig.~\ref{fig:mtt}). Recall that this
effect is a manifestation of spin correlations between the two steps of the
decay chain, and thus it is absent for a scalar $Y$. Furthermore, the 
invariant-mass distribution depends only the \emph{relative} chirality 
between the first and second interactions in the decay chain, that is, 
whether $\Gamma$ and $\Gamma'$ 
in Table~\ref{tab:models} have the same or opposite chirality.

Additional information on the couplings' chirality can be extracted from 
measurement of the top-quark polarization. The polarization can be
determined from the angular distribution of the top-quark decay products. This
method is particularly effective for large mass differences $m_{\cal Z}-m_Y$ or
$m_Y-m_X$, when the emitted top quarks are energetic, so their helicity is
approximate preserved.

Following the analysis in Ref.~\cite{cfhl}, we study the distribution of the
angle $\theta'_b$ of the $b$ quark with respect to the top-quark boost direction
in the top rest frame for the hadronically decaying top, $t_h$. Owing
to the left-handedness of the weak decay $t \to W^+b$, the $b$ quark is
emitted preferentially in the forward direction ($\cos\theta'_b>0$) if the
top quark is left-handed, and in the backward direction
($\cos\theta'_b<0$) if the top quark is right-handed.

\begin{figure}[t!]
\psfig{figure=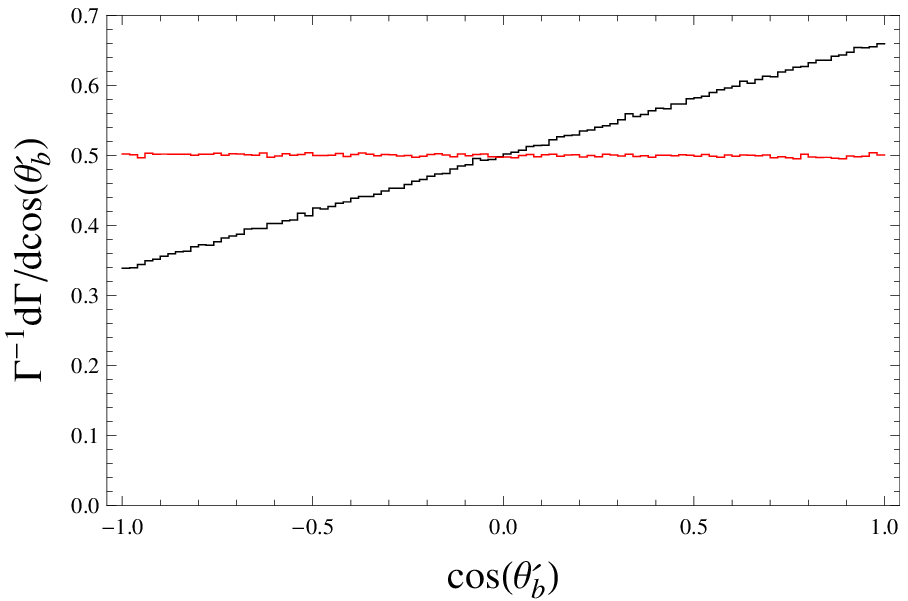, width=8.5cm, bb=0 25 260 173, clip=true}%
\rput[l](-3,1){scenario v+ii}\rput[l](-3,0.5){(SFS)}%
\psfig{figure=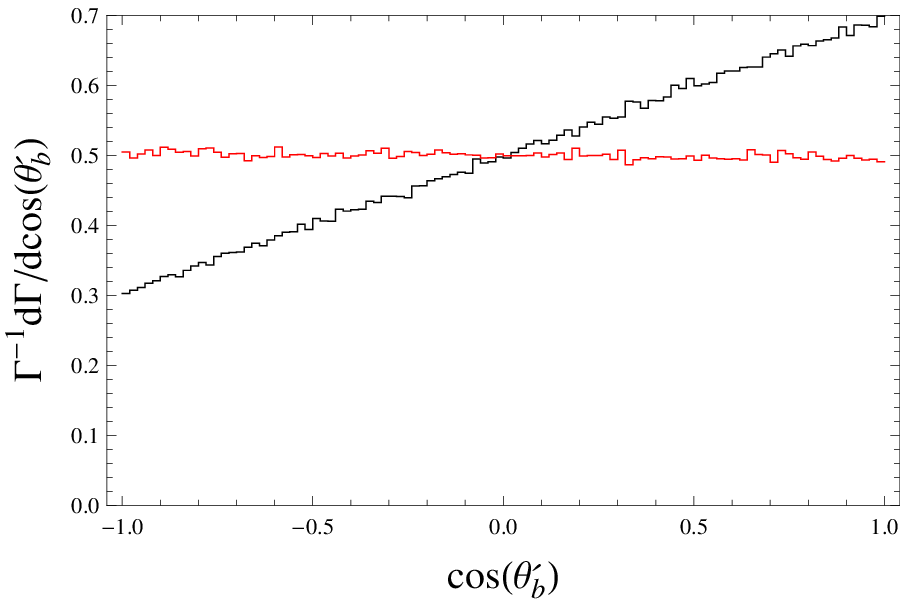, width=7.9cm, bb=18 25 260 173, clip=true}%
\rput[l](-3,1){scenario v+iii}\rput[l](-3,0.5){(SFV)}\\
\psfig{figure=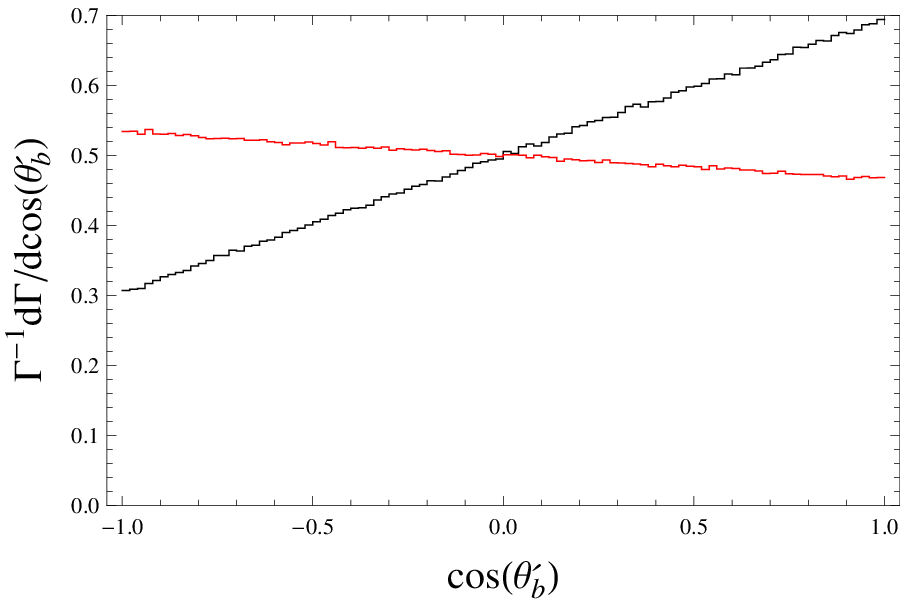, width=8.5cm, bb=0 25 260 173, clip=true}%
\rput[l](-3,1){scenario viii+ii}\rput[l](-3,0.5){(VFS)}%
\psfig{figure=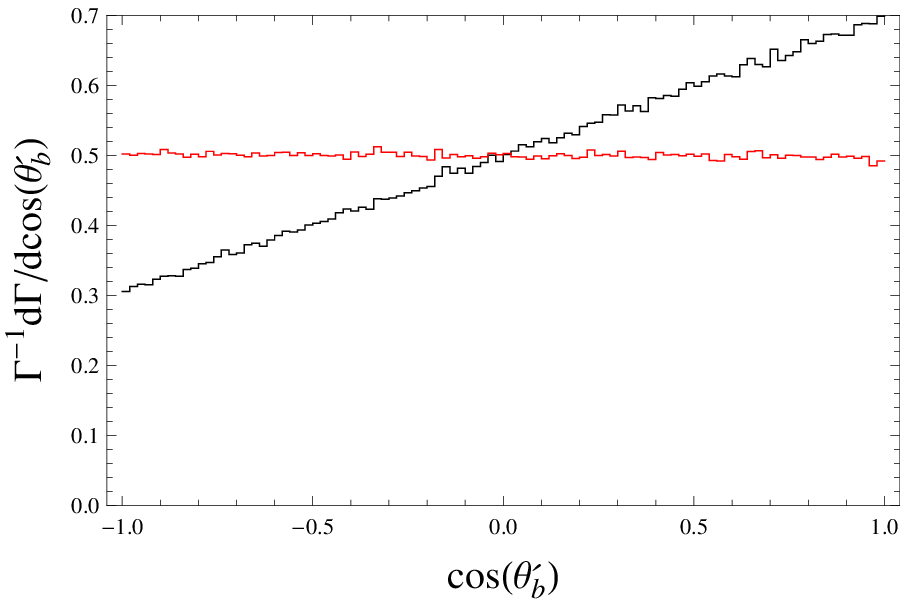, width=7.9cm, bb=18 25 260 173, clip=true}%
\rput[l](-3.1,1){scenario viii+iii}\rput[l](-3.1,0.5){(VFV)}\\
\psfig{figure=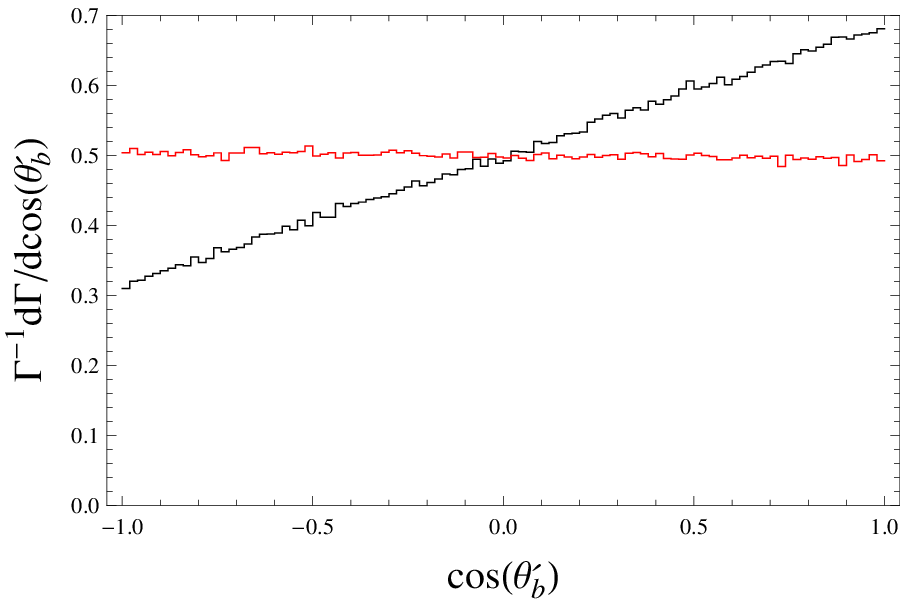, width=8.5cm}%
\rput[l](-3,1.8){scenario vii+iv}\rput[l](-3,1.3){(FVF)}%
\psfig{figure=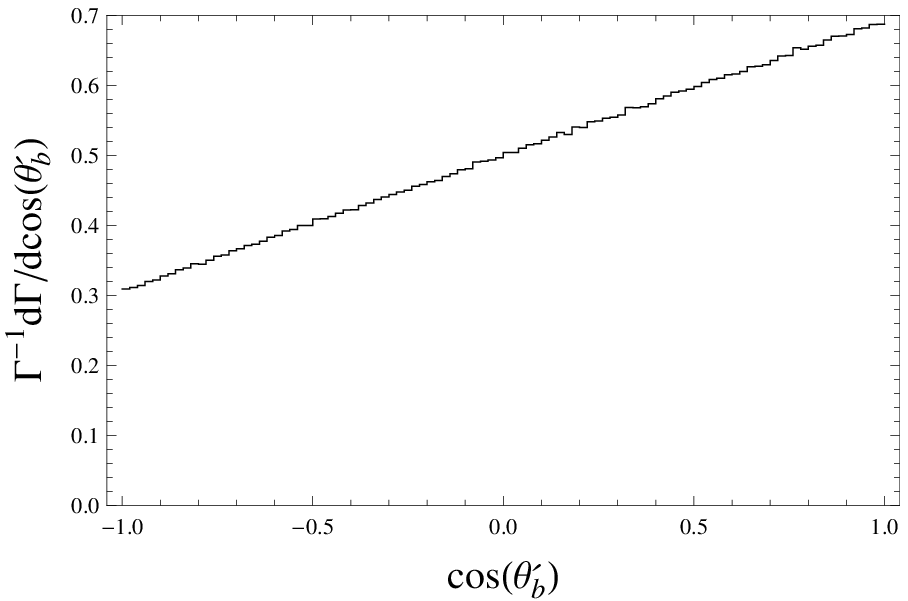, width=7.9cm, bb=18 0 260 173, clip=true}%
\rput[l](-3.1,1.8){scenario vi(a)+i}\rput[l](-3.1,1.3){(FSF)}%
\vspace{-.5em}
\mycaption{Parton-level angular distribution of the $b$-quark (jet) 
in the top-quark rest frame, for the decay chain 
${\cal Z} \to t\bar{Y}/\bar{t}Y \to t\bar{t}X$.
The six panels show the results for the scenarios in Table~\ref{tab:models}, 
for the two coupling choices $a_L=1,\,a_R=0,\,b_L=1,\,b_R=0$ (black) 
and $a_L=0,\,a_R=1,\,b_L=1,\,b_R=0$ (red). Here, S, F, and V denote 
scalar, fermion, and vector particles, respectively, in the decay chain. 
The input mass parameters are 
$m_{\cal Z}=1200\gev$, $m_Y=600\gev$ and $m_X=100\gev$. 
The distributions have been normalized to unity.
\label{fig:pol2}}
\end{figure}

The resulting $\cos\theta'_b$ distributions are shown in Fig.~\ref{fig:pol2},
based on a parton-level simulation with {\sc CalcHEP}. Since in general it is
unknown whether the $t_h$ emerged from the first or second step of the
decay chain, the observable distributions correspond to an average of both. 
Consequently, when $\Gamma$ and $\Gamma'$ (specified by $a_{L,R}$ and 
$b_{L,R}$) 
have the same chirality, the $\cos\theta'_b$ distribution displays a strong 
polarization signal. On the other hand, if they 
have opposite chirality, the top quarks from the two decay stages have
opposite polarization, leading to an almost flat average $\cos\theta'_b$
distribution. Thus, the polarization analysis allows one to determine the
chirality of the couplings regardless of the spins of $X$, $Y$ and $\cal Z$.

\subsection{Distinguishing between Majorana and Dirac Particles}

In general, the color-octet $\cal Z$ field may be self-conjugate or have
distinct particles and antiparticles. In this subsection, we investigate
whether these two possibilities can be distinguished experimentally at the LHC.
For concreteness, we focus on a color octet with spin $1/2$,
corresponding to a Majorana or Dirac gluino in supersymmetric theories. In
broad terms, there are two main approaches to distinguishing between 
self-conjugate and non-self-conjugate ${\cal Z}$ particles:
\paragraph{1.} One can take advantage of the fact that the production
cross section for the Dirac case is larger than for the Majorana case by a
factor of about 2; see Fig.~\ref{plot:prodz} and Ref.~\cite{Choi:2008pi}.
However, a difficulty associated with this method is that the total cross 
section also depends strongly on the spin of the ${\cal Z}$, its branching 
fractions and its mass, which is challenging to measure precisely.
\paragraph{2.} Alternatively, one can look for characteristics in the
decay distributions of the $\cal Z\bar{Z}$ pair. This is the approach we 
study in more detail here.

\vspace{\bigskipamount} In the non-self-conjugate (Dirac) case, the ${\cal Z}$
(gluino) decays into a  $Y$ (stop) and the ${\cal \bar{Z}}$ (anti-gluino) decays
into a $\bar{Y}$  (anti-stop).  For the same-sign lepton signature, this means
that one $\ell^+$ has to come from the second decay step of the ${\cal Z}$ and
the other from the first decay step of the ${\cal \bar{Z}}$ (and vice versa for
$\ell^-$). 
In the self-conjugate (Majorana) case, in contrast, the two decay chains
are independent, and therefore each of the two same-sign leptons can come  
from the same stage in its decay chain. 
Therefore, one expects the energy and $|{\bf p}_{T}|$ distributions of the 
two leptons to exhibit larger differences --- in particular, the $\ell_2$ 
will have a softer distribution --- in the Dirac case than in the 
Majorana case. 

The size of this effect depends crucially on kinematics. If $m_{\cal Z}-m_Y$
and $m_Y-m_X$ are approximately equal, the energy distributions of leptons
from the first and second decay steps differ very little, so the 
Majorana--Dirac distinction is difficult to make. On the
other hand, if $m_{\cal Z}-m_Y$ is significantly larger than $m_Y-m_X$, a 
lepton emitted from the second decay step $Y \to tX$ is on average softer then 
one from the first decay step ${\cal Z} \to \bar{t}Y$, so an attempted
discrimination between Majorana and Dirac gluinos is promising. Even in this 
case, however, the effect is relatively small, so a large integrated
luminosity will be needed for this analysis.

To study the effectiveness of this method, we have performed a Monte Carlo
simulation of Majorana and Dirac gluino production at the LHC. The simulation 
of $\cal ZZ$ (or $\cal Z\bar{Z}$) pair production with the full decay chain,
including top and $W$ decays, with exact matrix elements is very difficult and
requires large computational resources. Here, the following simplified approach
has been taken: parton-level events for $pp \to t\bar{t}t\bar{t}XX$ have been
generated with {\sc CalcHEP} and passed to {\sc Pythia} to perform the 
top-quark decays. This setup is computationally efficient but ignores the 
top-quark polarization. Therefore, we have to restrict ourselves to observables that are based only on
kinematical features, such as the energy and $|{\bf p}_{T}|$ distributions
proposed above.

As concrete examples, we have considered two choices for the mass spectrum:
\begin{align}
{\rm A}: \qquad &m_{\cal Z} = 1200\gev, \quad m_Y = 600\gev, \quad m_X = 400\gev, \\
{\rm B}: \qquad &m_{\cal Z} = 1200\gev, \quad m_Y = 1000\gev, \quad m_X =
400\gev.
\end{align}
In scenario A, $m_Y-m_X \ll m_{\cal Z}-m_Y$, whereas in scenario B,
$m_Y-m_X \gg m_{\cal Z}-m_Y$.
For the numerical analysis, the same cuts as in section~\ref{future} have been
applied. With the production cross section for Dirac $\cal Z\bar{Z}$ pairs as
the reference scenario, this choice produces an event yield of 16,200 for 
scenario A and 15,970 for scenario B at $\sqrt{s}=14\tev$, with an integrated 
luminosity of 3,000~fb$^{-1}$.

\begin{figure}[t!]
\begin{tabular}{@{}cc@{}}
A: $(m_{\cal Z},\,m_Y,\,m_X) = (1200,600,400)$ GeV &
B: $(m_{\cal Z},\,m_Y,\,m_X) = (1200,1000,400)$ GeV\\[.8ex]
\epsfig{figure=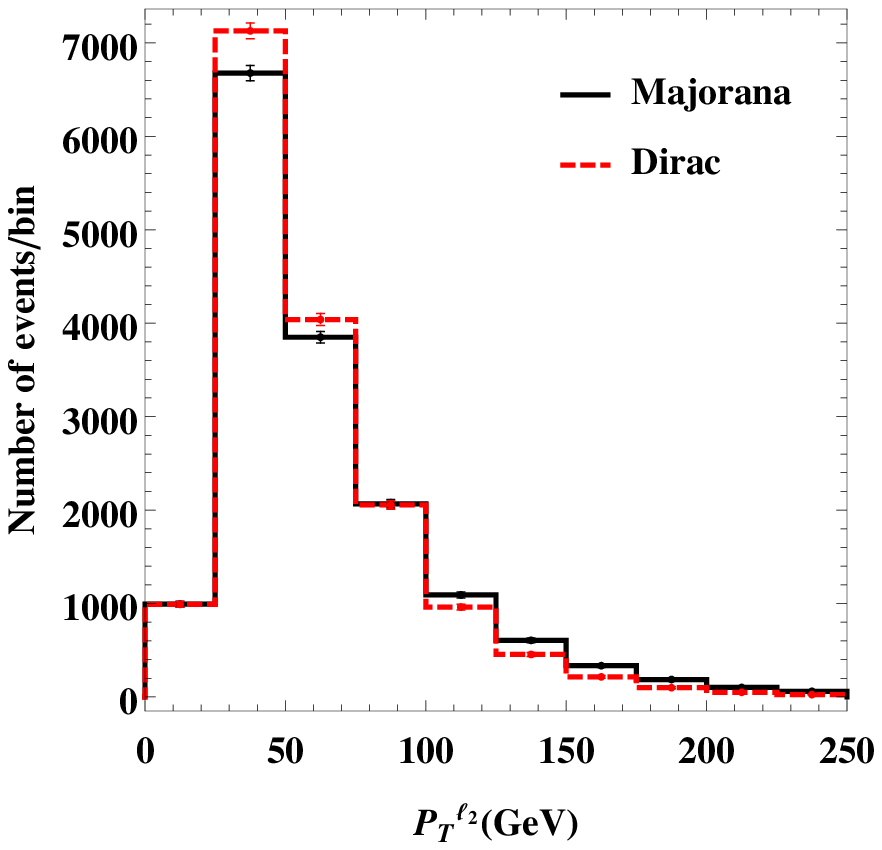, height=7.6cm} &
\epsfig{figure=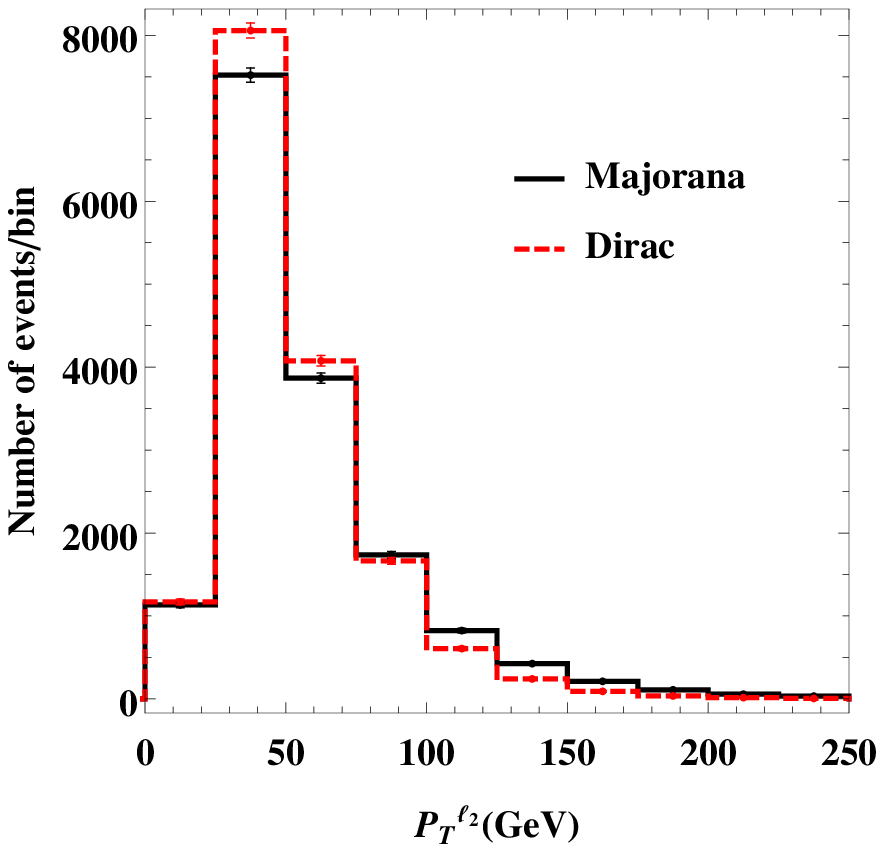, height=7.6cm} 
\\
\end{tabular}
\vspace{-1em}
\mycaption{Hadron-level 
$|\mbox{\bf p}_{T,\ell_2}|$ distribution from pair production of Majorana
fermion pairs $\cal ZZ$ (black solid) and Dirac fermion pairs $\cal Z\bar{Z}$ 
(red dashed). The Dirac and Majorana cases correspond to scenarios vi(a)+i and vi(b)+i in Table~\ref{tab:models}, respectively.
The chiral couplings have been fixed to $a_L=1,\,a_R=0,\,b_L=1,\,b_R=0$. The 
distributions have been normalized to the production cross section for Dirac 
octets after application of the selection cuts from section~\ref{future} at 
$\sqrt{s}=14\tev$ and $3000\ {\rm fb}^{-1}$ ($N_{\rm ev} = 16200$ and 
$N_{\rm ev} = 15970$ for scenarios A and B, respectively), and the error bars 
indicate the statistical uncertainty.
\label{fig:majdir}}
\end{figure}

The resulting $|\mbox{\bf p}_{T,\ell_2}|$ 
distributions are shown in Fig.~\ref{fig:majdir}. 
The softer $|\mbox{\bf p}_{T}|$ spectrum of the second lepton in
the Dirac case can be clearly seen in both scenarios.
Performing a binned $\chi^2$ analysis with three bins for each 
distribution,
one obtains the following levels of statistical discrimination between 
Majorana and Dirac octets:
\begin{align} 
&\text{300 fb}^{-1}\phantom{0}: &
{\rm A}: \quad &\sqrt{\chi^2} = 3.2, &
{\rm B}: \quad &\sqrt{\chi^2} = 4.1; \\
&\text{3000 fb}^{-1}: &
{\rm A}: \quad &\sqrt{\chi^2} = 10.1, &
{\rm B}: \quad &\sqrt{\chi^2} = 13.1.
\end{align}
Thus, a statistically significant exclusion of the 
scenario not realized in nature may be achievable
at the full-energy run of the LHC.


\section{Conclusions}
\label{concl}

In this paper, we have introduced a general categorization of new particles 
motivated by naturalness arguments, with different spin (0, 1/2, and 1) and 
color (octet $\cal Z$, triplet $Y$, and singlet $X$).
There are four possible spin combinations permitting an interaction
between the color triplet, singlet and top quark, 
and four possibilities for a coupling between the octet, triplet and top, 
as summarized in Table~\ref{tab:models}. 
The cross sections for the pair production of heavy color-octet particles, 
$\cal Z\bar{Z}$, at LHC energies are shown in Fig.~\ref{plot:prodz}. These 
channels would lead to a spectacular signature of four top quarks and
missing energy (see  Eq.~\eqref{eq:process}), where it is assumed that the 
singlet $X$ is stable and escapes detection.
At the 14-TeV run of the LHC, this process is observable 
at the $5\sigma$ level up to gluon-partner masses of 1280 (1480)~GeV for a
scalar ${\cal Z}$, 1650 (1860)~GeV for a fermionic ${\cal Z}$, and 1900 (2100)~GeV for 
a vector ${\cal Z}$ with an integrated luminosity of 300 (3000) fb$^{-1}$, 
provided that the missing particle $X$ (the dark-matter 
candidate) is not too heavy, that is, $m_X \lesssim 300\gev$.
These results are summarized in Table \ref{tab:reach}.

If such a signal is discovered, understanding the underlying physics 
will require 
the determination of properties of the new particles.
As a benchmark, we have taken the typical production rate of Majorana
color-octet fermions with ${\cal O}$(TeV) mass at $\sqrt{s}=14\tev$ 
with an integrated luminosity of $300 \, \textrm{fb}^{-1}$. Through an 
analysis of the invariant-mass distribution of the visible decay products
of top-quark pairs, most possible spin combinations can be discriminated from each other 
with high significance; 
$\sqrt{\chi^2}$ values are shown in Table~\ref{tab:spinchisq}. 
However, pairs that differ only in the spin of the invisible color singlet $X$ 
are difficult to distinguish.
 
Furthermore, as Fig.~\ref{fig:mtt} shows, the observable invariant-mass 
distribution is also affected by the chiral structure of the couplings of 
the particles in the decay chain, that is, whether they are left- or 
right-handed. Additional information on the couplings' chirality can be 
extracted from the top-quark polarization, which can be determined from 
the angular distribution of the top-quark decay products. 
The resulting $\cos\theta'_b$ distributions are shown in Fig.~\ref{fig:pol2}.
The polarization analysis allows one to determine the
chirality of the couplings independently of the spins of $X$, $Y$ and $\cal Z$.

Finally, for the case of fermionic color-octet pair production, we have 
demonstrated that measurements at the LHC also allow us to distinguish 
whether these particles are Majorana or Dirac fermions, without recourse to 
the factor of 2 difference in the production cross sections. 
This is possible because, for a pair of Majorana particles, each can 
decay randomly and independently into a top quark or antiquark, whereas 
fermion number is conserved in the decays of a Dirac fermion. 
Consequently, 
depending on the mass hierarchy of the new $\cal Z$, $Y$ and $X$ particles, 
there can be distinct differences in the transverse-momentum distributions 
of the final-state decay products, as seen in Fig.~\ref{fig:majdir}.

We have shown that the full-energy run of the LHC will have a significantly 
expanded potential for searching for heavy color octets and triplets, as 
well as identifying their characteristic properties, which could lead to a 
new understanding of the naturalness of the electroweak scale.
At a future 100-TeV VLHC, the mass coverage for a color-octet particle can 
be substantially extended, with a cross section many orders of magnitude 
greater than at the LHC, enabling the probing of ${\cal Z}$ masses of the 
order of 10 TeV.


\section*{Acknowledgements}

This project was supported in part by the National Science Foundation 
under grant PHY-1212635, by the US Department of Energy under grant 
Nos.~DE-FG02-12ER41832 and DE-AC02-98CH10886, and by PITT PACC.
Perimeter Institute is supported in part by the Government of Canada through 
Industry Canada and by the Province of Ontario through the Ministry of 
Research and Innovation.




\end{document}